\documentclass[11pt]{article}
\pdfoutput=1

\usepackage{bm,wrapfig,float,array,mathtools}
\usepackage{amsfonts}
\usepackage{amssymb}
\usepackage{cite}
\usepackage{amsmath}
\usepackage{color}
\usepackage{graphicx}
\usepackage{hyperref}
\usepackage{float}
\usepackage[T1]{fontenc}
\usepackage[utf8]{inputenc}
\usepackage{alphabeta}
\usepackage{fancyvrb}
\usepackage[dvipsnames]{xcolor}
\usepackage{bold-extra}
\usepackage{lmodern}
\usepackage{cleveref}
\usepackage[normalem]{ulem}
\usepackage{tikz-cd}
\newcommand{\RNum}[1]{\uppercase\expandafter{\romannumeral #1\relax}}

\graphicspath{ {figures/} }

\usepackage{tikz}
\usetikzlibrary{arrows}
\usetikzlibrary{arrows.meta}
\usetikzlibrary{shapes.geometric,calc,arrows, positioning,shapes.misc,decorations.markings}
\tikzset{
  big arrow/.style={
    decoration={markings,mark=at position 1 with {\arrow[scale=2,#1]{>}}},
    postaction={decorate},
    shorten >=0.4pt},
  big arrow/.default=black}
  
\pgfdeclarelayer{edgelayer}
\pgfdeclarelayer{nodelayer}
\pgfsetlayers{edgelayer,nodelayer,main} 
\tikzstyle{none}=[inner sep=0pt] 

\tikzstyle{NodeCross}=[draw, shape=circle, cross out, inner sep=0pt, minimum size=6pt,line width=0.25mm]
\tikzstyle{Circle}=[draw, shape=circle, black, inner sep=0pt, minimum size=6pt]
\tikzstyle{rtriangle}=[fill=black, regular polygon, regular polygon sides=3, rotate=90, inner sep=0pt, minimum size=8pt]
\tikzstyle{ltriangle}=[fill=black, regular polygon, regular polygon sides=3, rotate=270, inner sep=0pt, minimum size=8pt]
\tikzstyle{rtriangleblue}=[fill={rgb,255: red,17; green,160; blue,255}, regular polygon, regular polygon sides=3, rotate=90, inner sep=0pt, minimum size=8pt]
\tikzstyle{ltriangleblue}=[fill={rgb,255: red,17; green,160; blue,255}, regular polygon, regular polygon sides=3, rotate=270, inner sep=0pt, minimum size=8pt]
\tikzstyle{rtrianglegreen}=[fill={rgb,255: red,69; green,255; blue,28}, regular polygon, regular polygon sides=3, rotate=90, inner sep=0pt, minimum size=8pt]
\tikzstyle{ltrianglegreen}=[fill={rgb,255: red,69; green,255; blue,28}, regular polygon, regular polygon sides=3, rotate=270, inner sep=0pt, minimum size=8pt]
\tikzstyle{Uprtriangle}=[fill=black, regular polygon, regular polygon sides=3, rotate=0, inner sep=0pt, minimum size=8pt]
\tikzstyle{Downltriangle}=[fill=black, regular polygon, regular polygon sides=3, rotate=180, inner sep=0pt, minimum size=8pt]
\tikzstyle{rtriangleAmber}=[fill={rgb,255: red, 191; green, 144; blue, 63}, regular polygon, regular polygon sides=3, rotate=90, inner sep=0pt, minimum size=8pt]
\tikzstyle{UprtriangleViolett}=[fill={rgb,255: red,255; green,0; blue,0}, regular polygon, regular polygon sides=3, rotate=0, inner sep=0pt, minimum size=8pt]
\tikzstyle{ltrianglered}=[fill={rgb,255: red,191; green,0; blue,0}, regular polygon, regular polygon sides=3, rotate=270, inner sep=0pt, minimum size=8pt]

\tikzstyle{Downltriangle}=[fill=black, regular polygon, regular polygon sides=3, rotate=180, inner sep=0pt, minimum size=8pt]
\tikzstyle{UpRighttriangle}=[fill=black, regular polygon, regular polygon sides=3, rotate=45, inner sep=0pt, minimum size=8pt]
\tikzstyle{UpLefttriangle}=[fill=black, regular polygon, regular polygon sides=3, rotate=315, inner sep=0pt, minimum size=8pt]
\tikzstyle{DownRighttriangle}=[fill=black, regular polygon, regular polygon sides=3, rotate=135, inner sep=0pt, minimum size=8pt]
\tikzstyle{DownLighttriangle}=[fill=black, regular polygon, regular polygon sides=3, rotate=225, inner sep=0pt, minimum size=8pt]

\tikzstyle{Star}=[draw, shape=star, fill=black, star points=8, inner sep=0pt, minimum size=8pt]

\tikzstyle{DashedLine}=[-, densely dashed, line width=0.25mm]
\tikzstyle{DashedLineBrown}=[-, densely dashed, line width=0.25mm, draw={rgb,255: red,155; green,103; blue,51}]
\tikzstyle{DashedLineFall}=[-, densely dashed, line width=0.25mm, draw={rgb,255: red,195; green,0; blue,0}]
\tikzstyle{DashedLineViolett}=[-, densely dashed, line width=0.25mm, draw={rgb,255: red,139; green,41; blue,148}]
\tikzstyle{DottedLine}=[-, dotted, line width=0.25mm]
\tikzstyle{BlueLine}=[-, fill=none, draw={rgb,255: red,17; green,160; blue,255}, line width=0.25mm]
\tikzstyle{GreenLine}=[-, fill=none, draw={rgb,255: red,69; green,255; blue,28}, line width=0.25mm]
\tikzstyle{RedLine}=[-, draw={rgb,255: red,191; green,0; blue,0}, fill=none, line width=0.25mm]
\tikzstyle{LBRedLine}=[-, draw={rgb,255: red,255; green,0; blue,0}, fill=none, line width=0.25mm]
\tikzstyle{DashedLineRed}=[-, densely dashed, fill=none, draw={rgb,255: red,191; green,0; blue,0}, line width=0.25mm]
\tikzstyle{ThickLine}=[-, line width=0.25mm]
\tikzstyle{ViolettLine}=[-, draw={rgb,255: red,132; green,60; blue,191}, fill=none, line width=0.25mm]
\tikzstyle{ViolettDashedLine}=[-, densely dashed, draw={rgb,255: red,132; green,60; blue,191}, fill=none, line width=0.25mm]
\tikzstyle{AmberLine}=[-, draw={rgb,255: red,191; green,144; blue,63}, fill=none, line width=0.25mm]
\tikzstyle{DashedRedThick}=[-, densely dashed, fill=none, draw={rgb,255: red,191; green,0; blue,0}, line width=0.40mm]
\tikzstyle{DashedBlueThick}=[-, densely dashed, fill=none, black, line width=0.40mm]
\tikzstyle{DottedLineRed}=[-, dotted, line width=0.25mm, draw={rgb,255: red,191; green,0; blue,0}]
\tikzstyle{DottedLineBlue}=[-, dotted, line width=0.25mm, draw={rgb,255: red,17; green,160; blue,255}]
\tikzstyle{ArrowLineRight}=[-, -{Stealth[scale=1.75]}, line width=0.1mm, scale=5]
\tikzstyle{ArrowLineLeft}=[-, {Stealth[scale=1.75]}-, line width=0.1mm, scale=5]
\tikzstyle{ArrowLineRightBlue}=[-, -{Stealth[scale=1.75]}, line width=0.1mm, draw={rgb,255: red,17; green,160; blue,255}]
\tikzstyle{ArrowLineLeftBlue}=[-, {Stealth[scale=1.75]}-, line width=0.1mm, draw={rgb,255: red,17; green,160; blue,255}]
\tikzstyle{ArrowLineRightRed}=[-, -{Stealth[scale=1.75]}, line width=0.1mm, draw={rgb,255: red,191; green,0; blue,0}]
\tikzstyle{ArrowLineLeftRed}=[-, {Stealth[scale=1.75]}-, line width=0.1mm, draw={rgb,255: red,191; green,0; blue,0}]

\newcommand{\bea}{\begin{eqnarray}}
\newcommand{\eea}{\end{eqnarray}}
\newcommand{\be}{\begin{equation}}
\newcommand{\ee}{\end{equation}}
\newcommand{\ba}{\begin{aligned}}
\newcommand{\ea}{\end{aligned}}
\newcommand{\bit}{\begin{itemize}}
\newcommand{\eit}{\end{itemize}}
\newcommand{\ben}{\begin{enumerate}}
\newcommand{\een}{\end{enumerate}}

\renewcommand{\ni}{\noindent}

\newcommand{\wt}{\widetilde}
\newcommand{\wh}{\widehat}

\newcommand{\Z}{{\mathbb Z}}
\newcommand{\R}{{\mathbb R}}

\newcommand{\Aut}{\text{Aut}}
\newcommand{\Spin}{\text{Spin}}
\newcommand{\Sp}{\text{Sp}}

\newcommand{\cA}{\mathcal{A}}

\newcommand{\cD}{\mathcal{D}}
\newcommand{\cE}{\mathcal{E}}
\newcommand{\cF}{\mathcal{F}}
\newcommand{\cG}{\mathcal{G}}

\newcommand{\cM}{\mathcal{M}}
\newcommand{\cN}{\mathcal{N}}
\newcommand{\cO}{\mathcal{O}}

\newcommand{\cS}{\mathcal{S}}

\newcommand{\cZ}{\mathcal{Z}}
\newcommand{\Bock}{\text{Bock}}

\newcommand{\F}{\mathsf{F}}

\renewcommand{\L}{\mathsf{\Lambda}}

\newcommand{\fG}{\mathfrak{G}}
\newcommand{\fT}{\mathfrak{T}}

\newcommand{\ff}{\mathfrak{f}}
\newcommand{\fg}{\mathfrak{g}}

\newcommand{\su}{\mathfrak{su}}
\renewcommand{\sp}{\mathfrak{sp}}
\newcommand{\so}{\mathfrak{so}}
\renewcommand{\u}{\mathfrak{u}}

\makeatletter 
\newcommand\newsize{\@setfontsize\newsize{23pt}{18}}
\makeatother
\begin{document}

\baselineskip=18pt  
\numberwithin{equation}{section}  
\allowdisplaybreaks  

\thispagestyle{empty}

%
\vspace*{0.8cm} 
\begin{center}
{{\newsize Disconnected 0-Form and 2-Group Symmetries}}

 \vspace*{1.5cm}
{\large Lakshya Bhardwaj, Dewi S.W. Gould}\\

 \vspace*{.2cm} 
{\it  Mathematical Institute, University of Oxford, \\
Andrew-Wiles Building,  Woodstock Road, Oxford, OX2 6GG, UK}

\vspace*{1.5cm}
\end{center}
\vspace*{1cm}

\noindent
Quantum field theories can have both continuous and finite 0-form symmetries. We study global symmetry structures that arise when both kinds of 0-form symmetries are present. The global structure associated to continuous 0-form symmetries is described by a connected Lie group, which captures the possible backgrounds of the continuous 0-form symmetries the theory can be coupled to. Finite 0-form symmetries can act as outer-automorphisms of this connected Lie group. Consequently, possible background couplings to both continuous and finite 0-form symmetries are described by a disconnected Lie group, and we call the resulting symmetry structure a disconnected 0-form symmetry. Additionally, finite 0-form symmetries may act on the 1-form symmetry group. The 1-form symmetries and continuous 0-form symmetries may combine to form a 2-group, which when combined with finite 0-form symmetries leads to another type of 2-group, that we call a disconnected 2-group and the resulting symmetry structure a disconnected 2-group symmetry. Examples of arbitrarily complex disconnected 0-form and 2-group symmetries in any spacetime dimension are furnished by gauge theories: with 1-form symmetries arising from the center of the gauge group, continuous 0-form symmetries arising as flavor symmetries acting on matter content, and finite 0-form symmetries arising from outer-automorphisms of gauge and flavor Lie algebras.

\newpage

\tableofcontents

\newpage 

\section{Introduction}
Recent years have seen the exploration of many different types of generalized global symmetry structures in quantum field theories. The first type of generalized global symmetries to have been systematically studied are higher-form symmetries that were defined in the seminal paper \cite{Gaiotto:2014kfa}. After a few years, it was realized that higher-form symmetries of different degrees can interact with each other leading to more general symmetry structures, known as higher-group symmetries \cite{Sharpe:2015mja,Tachikawa:2017gyf}. This paper studies generalized global symmetries known as 2-group symmetries, which describe the interplay between 0-form and 1-form symmetries. 

Various types of 2-group symmetries were studied in initial works \cite{Cordova:2018cvg,Benini:2018reh,Hsin:2020nts,Cordova:2020tij}. More recently, systematic understanding was developed in \cite{Apruzzi:2021vcu,Bhardwaj:2021ojs,Bhardwaj:2021wif,Lee:2021crt} for a special type of 2-group symmetries, that we call Bockstein 2-group symmetries, involving finite 1-form symmetries and \emph{continuous} 0-form symmetries. It was shown in \cite{Bhardwaj:2021wif}, building on the work of \cite{Hsin:2020nts}, that such Bockstein 2-group symmetries are ubiquitous in gauge theories in arbitrary spacetime dimension. This systematic understanding also paved the way for determining Bockstein 2-group symmetries in non-Lagrangian theories like 4d $\cN=2$ Class S theories \cite{Bhardwaj:2021wif}, and 5d and 6d SCFTs \cite{Apruzzi:2021vcu,Apruzzi:2021mlh,DelZotto:2022joo,Cvetic:2022imb}.\footnote{See \cite{Bhardwaj:2021pfz,Nguyen:2021naa,Heidenreich:2021xpr,Apruzzi:2021phx,Hosseini:2021ged,Cvetic:2021sxm,Buican:2021xhs,Bhardwaj:2021zrt,Iqbal:2021rkn,Braun:2021sex,Cvetic:2021maf,Closset:2021lhd,Hidaka:2021mml,Lee:2021obi,Hidaka:2021kkf,Koide:2021zxj,Kaidi:2021xfk,Choi:2021kmx,Bah:2021brs,Gukov:2021swm,Closset:2021lwy,Yu:2021zmu,Apruzzi:2021nmk,Beratto:2021xmn,Bhardwaj:2021mzl,DelZotto:2022fnw,Chatterjee:2022kxb,Benini:2022hzx,Apruzzi:2022dlm,Hubner:2022kxr,Lee:2022spd,Carta:2022spy,Roumpedakis:2022aik,Bhardwaj:2022yxj,DelZotto:2022ras,Hayashi:2022fkw,Choi:2022zal,Argyres:2022kon,Kaidi:2022uux,Heckman:2022suy,Benedetti:2022zbb,Choi:2022jqy,Cordova:2022ieu,Chatterjee:2022tyg,Lohitsiri:2022jyz,Pantev:2022kpl,Sharpe:2021srf,Robbins:2021xce,Bhardwaj:2022dyt,Bolognesi:2022beq} for other recent work on generalized global symmetries.}

In this work, we generalize the considerations of \cite{Bhardwaj:2021wif} to include \emph{finite} 0-form symmetries along with the continuous 0-form symmetries considered there. A new ingredient that arises is that the finite 0-form symmetry can now act on both the 1-form symmetries and the continuous 0-form symmetries. One of the consequences of this action is that the total 0-form symmetry group, incorporating both continuous and finite 0-form symmetries, becomes a disconnected Lie group\footnote{Disconnected gauge groups, as opposed to disconnected 0-form global symmetry groups, have been discussed in a number of recent works, see \cite{Argyres:2016yzz,Cordova:2017vab,Bourget:2018ond,Arias-Tamargo:2019jyh,Arias-Tamargo:2021ppf,Henning:2021ctv}.} which is expressed as a semi-direct product of continuous and finite parts of the 0-form symmetry.

Moreover, the action of finite 0-form on 1-form and continuous 0-form symmetries combines to form an action of finite 0-form on the 2-group symmetry formed by the 1-form and continuous 0-form symmetries. Correspondingly, the total 2-group structure, including the finite 0-form symmetry, can be thought of as a \textit{disconnected 2-group} which is morally a semi-direct product of the `connected' 2-group symmetry discussed above with the finite 0-form symmetry. Correspondingly, the 2-group symmetry is modified to what we call a \textit{disconnected 2-group symmetry}.

We describe the general structure of a disconnected 2-group symmetry and discuss in detail a special class of disconnected 2-group symmetries that we call disconnected Bockstein 2-group symmetries. A key role in the definition of a disconnected Bockstein 2-group symmetry is played by a generalization of Bockstein homomorphism, that we call \textit{twisted Bockstein} homomorphism, which is the connecting homomorphism in long exact sequence of \textit{twisted cohomology} groups associated to a short exact sequence with an action of a finite group on the short exact sequence. 

We show that disconnected Bockstein 2-group symmetries arise naturally in gauge theories in any spacetime dimension, and develop methods to compute such disconnected 2-group symmetries. We hope that, as in the case for connected Bockstein 2-group symmetries, the understanding of disconnected Bockstein 2-group symmetries in gauge theories will pave the way for determining disconnected 2-group symmetries in non-Lagrangian 4d and higher dimensional theories. We will report on this in future work.

One interesting point to note is that a disconnected 2-group can be non-trivial even if its connected component 2-group is trivial. This is discussed at the end of subsection \ref{sec:discconn2groupsym} and an explicit gauge theory example is provided around equation (\ref{CdC}).

Finally, let us remark that a disconnected 2-group symmetry is naturally associated to a mixed 't Hooft anomaly of higher-form symmetries, with the two being related by gauging of 1-form symmetry. This provides a way for constructing 't Hooft anomalies, e.g. anomalies involving characteristic classes of degree 1 of a disconnected Lie group, like first Stiefel-Whitney class of the orthogonal group $O(n)$.

The structure of this paper is as follows:
\bit
\item In section \ref{sec:0actingon1}, we discuss the action of finite 0-form symmetries on 1-form symmetries, and show that natural examples of such situations in arbitrary spacetime dimension are provided by 0-form symmetries descending from outer-automorphisms of gauge and flavor Lie algebras in a gauge theory.
\item In section \ref{sec:disconnected0fs}, we discuss the action of finite 0-form symmetries on continuous 0-form symmetries. Due to this action, the combined group structure of the full 0-form symmetry, incorporating both continuous and finite parts, is described by a disconnected Lie group. We discuss in detail how disconnected 0-form symmetry groups can be determined in gauge theories in arbitrary spacetime dimension.
\item In section \ref{sec:discon2groups}, we discuss the action of finite 0-form symmetries on 2-group symmetries involving finite 1-form and continuous 0-form symmetries. Due to this action, the combined symmetry structure becomes another type of 2-group, that we call a disconnected 2-group. We discuss in detail a special type of disconnected 2-group symmetries that we refer to as disconnected Bockstein 2-group symmetries, and describe the determination of such 2-group symmetries in gauge theories in arbitrary spacetime dimension. We also introduce the notion of a disconnected structure group whose bundles describe the gauge and flavor bundles that can appear for a fixed background of a disconnected Bockstein 2-group symmetry. At the end of the section, we describe a mixed 't Hooft anomaly of higher-form symmetries which arises by gauging the 1-form symmetry part of a disconnected 2-group.
\item Finally, we have some appendices. In appendix \ref{sec:liealgeouterreview} we review outer-automorphisms of Lie algebras and how they arise from symmetries of the associated Dynkin diagram. In appendix \ref{sec:semidirprod} we review semi-direct products which are at the heart of the construction of the disconnected Lie groups considered in this work. In appendix \ref{sec:twistedcoho} we review twisted cohomology and use that machinery in appendix \ref{sec:bockrev} to introduce the twisted Bockstein homomorphism which plays a key role in the disconnected Bockstein 2-groups we consider.
\eit
All the considerations of this paper are illustrated through a variety of gauge theory examples.

\section{0-Form Symmetries Acting on 1-Form Symmetries}
\label{sec:0actingon1}
\subsection{General Theory}
We begin this paper with the discussion of an abstract scenario in which we have a finite 0-form symmetry group $\Gamma^{(0)}$ (which can be non-abelian) whose elements can act on the 1-form symmetry group $\Gamma^{(1)}$ (which must be abelian, but can be finite or continuous). In later subsections, we describe how outer-automorphisms of continuous gauge and 0-form symmetries provide natural examples of such a scenario.

\begin{figure}
\centering
\begin{tikzpicture}
\draw[fill=red,opacity=0.5] (-1,1) -- (-1,2.5) -- (1,1) -- (1,-3) -- (-1,-1.5)--(-1,1);
\draw[thick,blue] (-4.5,-0.5) -- (0,-0.5);
\draw [thick,brown] (4.5,-0.5) -- (1,-0.5);
\draw [brown,dashed](0,-0.5) -- (1,-0.5);
\node[blue] at (-5.5,-0.5) {$\gamma\in\Gamma^{(1)}$};
\node[brown] at (6,-0.5) {$o\cdot\gamma\in\Gamma^{(1)}$};
\node[red] at (1,2) {$o\in\Gamma^{(0)}$};
\draw [black,fill=black] (0,-0.5) ellipse (0.05 and 0.05);
\begin{scope}[shift={(1.5,-7)}]
\draw[fill=red,opacity=0.5] (0.5,1) -- (0.5,2.5) -- (2.5,1) -- (2.5,-3) -- (0.5,-1.5)--(0.5,1);
\begin{scope}[shift={(-3.5,0)}]
\draw[fill=red,opacity=0.5] (-1,1) -- (-1,2.5) -- (1,1) -- (1,-3) -- (-1,-1.5)--(-1,1);
\end{scope}
\draw[thick,blue] (-6,-2) -- (-6,1.5);
\draw [thick,brown] (4,-2) -- (4,1.5);
\node[blue] at (-6,-2.5) {$\gamma\in\Gamma^{(1)}$};
\node[brown] at (4,-2.5) {$o\cdot\gamma\in\Gamma^{(1)}$};
\node[red] at (2.5,2) {$o\in\Gamma^{(0)}$};
\node at (-1,-0.5) {=};
\node[red] at (-2.5,2) {$o\in\Gamma^{(0)}$};
\end{scope}
\end{tikzpicture}
\caption{Two ways of understanding the action of $\Gamma^{(0)}$ on $\Gamma^{(1)}$. Top: A codimension-two topological defect $\gamma\in\Gamma^{(1)}$ travelling through a codimension-one topological defect $o\in\Gamma^{(0)}$ emerges as another codimension-two topological defect $o\cdot\gamma\in\Gamma^{(1)}$. Bottom: Passing a codimension-one topological defect $o\in\Gamma^{(0)}$ across a codimension-two topological defect $\gamma\in\Gamma^{(1)}$ changes it into another codimension-two topological defect $o\cdot\gamma\in\Gamma^{(1)}$.}
\label{fig:zeroactingonone}
\end{figure}
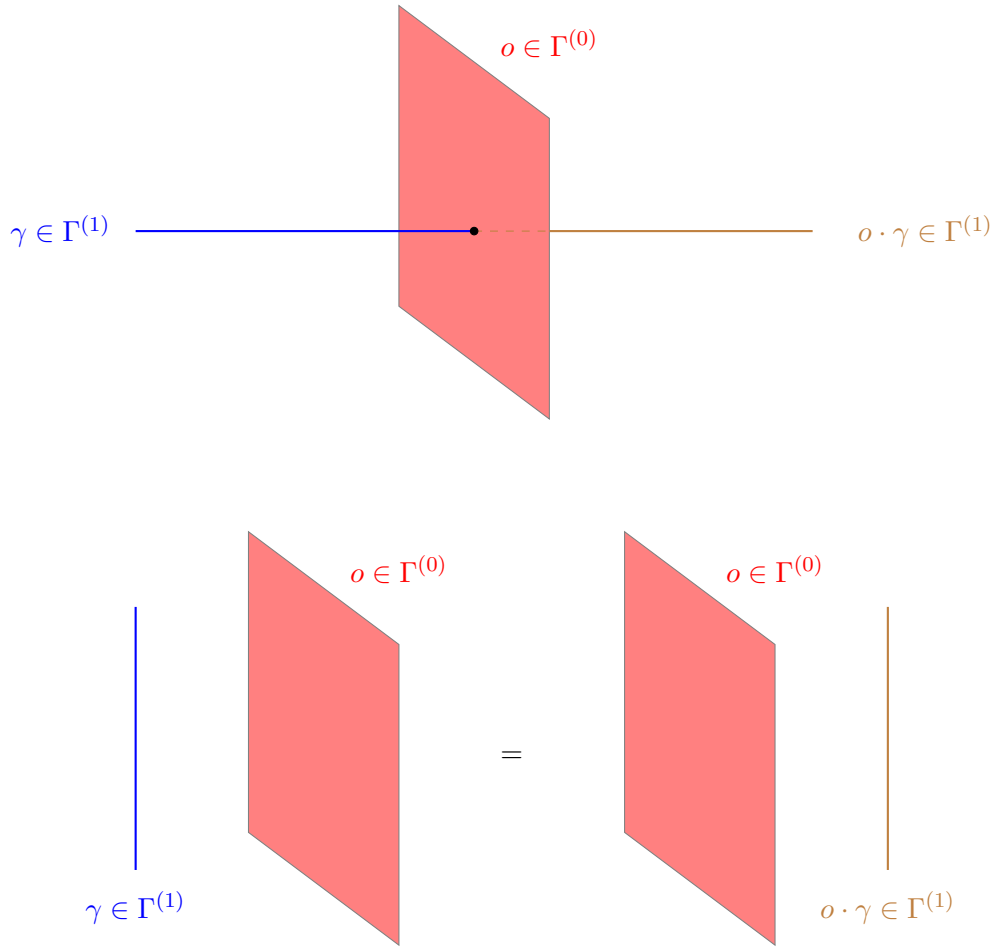

\paragraph{Action of 0-form on 1-form.}
The action of $\Gamma^{(0)}$ on $\Gamma^{(1)}$ has the following physical interpretation. Let us consider a codimension-two topological defect generating a 1-form symmetry $\gamma\in\Gamma^{(1)}$ and transport it through a codimension-one topological defect generating a 0-form symmetry $o\in\Gamma^{(0)}$. Then, the codimension-two topological defect generating $\gamma$ is converted into another codimension-two topological defect generating $o\cdot\gamma\in\Gamma^{(1)}$, which is the element of $\Gamma^{(1)}$ obtained by applying the action of $o$ on $\gamma$ (figure \ref{fig:zeroactingonone}).

\begin{figure}
\centering
\begin{tikzpicture}
\draw[fill=red,opacity=0.5] (-1,1) -- (-1,2.5) -- (1,1) -- (1,-3) -- (-1,-1.5)--(-1,1);
\draw[thick] (-4.5,-0.5) -- (0,-0.5);
\draw [thick,teal] (4.5,-0.5) -- (1,-0.5);
\draw [teal,dashed](0,-0.5) -- (1,-0.5);
\node[] at (-5.5,-0.5) {$\wh\gamma\in\wh\Gamma^{(1)}$};
\node[teal] at (6,-0.5) {$o\cdot\wh\gamma\in\wh\Gamma^{(1)}$};
\node[red] at (1,2) {$o\in\Gamma^{(0)}$};
\draw [black,fill=black] (0,-0.5) ellipse (0.05 and 0.05);
\draw [ultra thick,white](-2.3,-0.5) -- (-2.1,-0.5);
\draw [thick,blue] (-2.5,-0.5) ellipse (0.3 and 0.7);
\draw [ultra thick,white](-2.8,-0.6) -- (-2.8,-0.4);
\draw [dashed,thick](-2.8396,-0.5) -- (-2.7396,-0.5);
\node[blue] at (-2.5,0.5) {$o^{-1}\cdot\gamma\in\Gamma^{(1)}$};
\begin{scope}[shift={(0,-5.5)}]
\draw[fill=red,opacity=0.5] (-1,1) -- (-1,2.5) -- (1,1) -- (1,-3) -- (-1,-1.5)--(-1,1);
\draw[thick] (-4.5,-0.5) -- (0,-0.5);
\draw [thick,teal] (4.5,-0.5) -- (1,-0.5);
\draw [teal,dashed](0,-0.5) -- (1,-0.5);
\node[] at (-5.5,-0.5) {$\wh\gamma\in\wh\Gamma^{(1)}$};
\node[teal] at (6,-0.5) {$o\cdot\wh\gamma\in\wh\Gamma^{(1)}$};
\node[red] at (1,2) {$o\in\Gamma^{(0)}$};
\draw [black,fill=black] (0,-0.5) ellipse (0.05 and 0.05);
\end{scope}
\begin{scope}[shift={(5,-5.5)}]
\draw [ultra thick,white](-2.3,-0.5) -- (-2.1,-0.5);
\draw [thick,brown] (-2.5,-0.5) ellipse (0.3 and 0.7);
\draw [ultra thick,white](-2.8,-0.6) -- (-2.8,-0.4);
\draw [dashed,thick,teal](-2.8396,-0.5) -- (-2.7396,-0.5);
\node[brown] at (-2.5,0.5) {$\gamma\in\Gamma^{(1)}$};
\end{scope}
\begin{scope}[shift={(0,-11)}]
\draw[fill=red,opacity=0.5] (-1,1) -- (-1,2.5) -- (1,1) -- (1,-3) -- (-1,-1.5)--(-1,1);
\draw[thick] (-4.5,-0.5) -- (0,-0.5);
\draw [thick,teal] (4.5,-0.5) -- (1,-0.5);
\draw [teal,dashed](0,-0.5) -- (1,-0.5);
\node[] at (-5.5,-0.5) {$\wh\gamma\in\wh\Gamma^{(1)}$};
\node[teal] at (6,-0.5) {$o\cdot\wh\gamma\in\wh\Gamma^{(1)}$};
\node[red] at (1,2) {$o\in\Gamma^{(0)}$};
\draw [black,fill=black] (0,-0.5) ellipse (0.05 and 0.05);
\end{scope}
\node at (-5.5,2.5) {(1)};
\node at (-5.5,-3) {(2)};
\node at (-5.5,-8.5) {(3)};
\end{tikzpicture}
\caption{All three correlation functions shown in the figure involve a line defect with charge $\wh\gamma\in\Gamma^{(1)}$ passing through a topological codimension-one defect $o\in\Gamma^{(0)}$, which converts it into another line defect with charge $o\cdot\wh\gamma\in\Gamma^{(1)}$. In addition, the correlation function (1) involves a topological codimension-two defect $o^{-1}\cdot\gamma\in\Gamma^{(1)}$ linking the line defect with charge $\wh\gamma$, and the correlation function (2) involves a topological codimension-two defect $\gamma\in\Gamma^{(1)}$ linking the line defect with charge $o\cdot\wh\gamma$. The correlations functions (1) and (2) are the same as we discussed above. Now the correlation function (1) can be related to the correlation function (3) by contracting $o^{-1}\cdot\gamma$ on top of $\wh\gamma$, leading to an additional phase factor $\wh\gamma(o^{-1}\gamma)\in U(1)$. Similarly, the correlation function (2) can be related to the correlation function (3) by contracting $\gamma$ on top of $o\cdot\wh\gamma$, leading to an additional phase factor $o\cdot\wh\gamma(\gamma)\in U(1)$. Consistency leads to the equation (\ref{eq:chargeconsistency}).}
\label{fig:dualaction}
\end{figure}
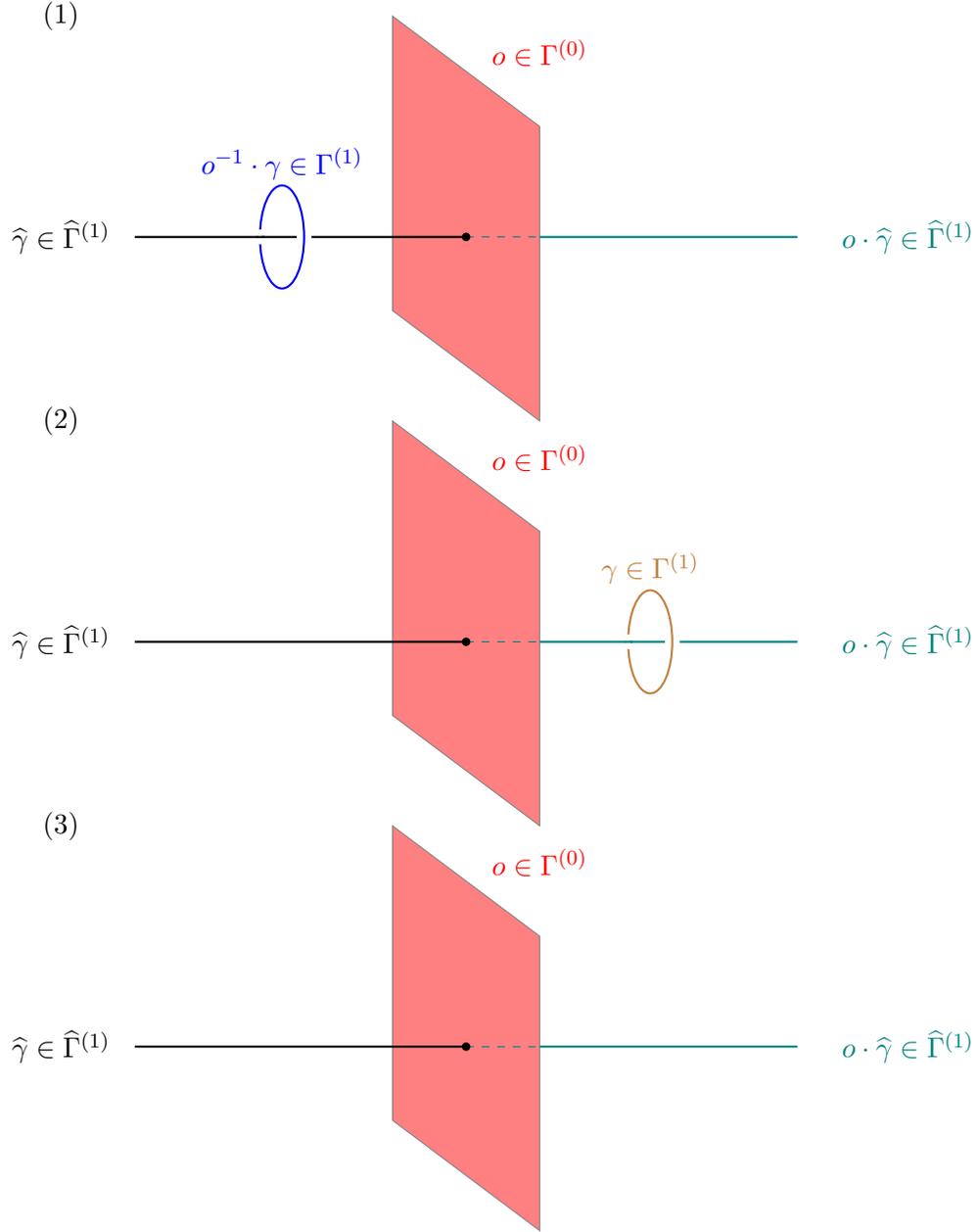

\paragraph{Dual action of 0-form on line defects.}
We also have a dual action of $\Gamma^{(0)}$ on the Pontryagin dual group
\be
\wh\Gamma^{(1)}:=\text{Hom}\left(\Gamma^{(1)},U(1)\right)\,,
\ee
which characterises the charges of the genuine line defects charged under $\Gamma^{(1)}$. This action is obtained as follows. Let $\wh\gamma\in\wh\Gamma^{(1)}$ and $o\in\Gamma^{(0)}$ be arbitrary elements. Then $o\cdot\wh\gamma$ is defined as the element that satisfies the condition
\be\label{eq:chargeconsistency}
o\cdot\wh\gamma\,(\gamma)=\wh\gamma\left(o^{-1}\cdot\gamma\right)\in U(1)\,,
\ee
for all $\gamma\in\Gamma^{(1)}$. This action has similar physical interpretation as for the action on $\Gamma^{(1)}$ discussed in figure \ref{fig:zeroactingonone}. We can justify the definition of the dual action using these physical interpretations, as explained in figure \ref{fig:dualaction}.

\paragraph{Backgrounds.}
If there is a non-trivial action of $\Gamma^{(0)}$ on $\Gamma^{(1)}$, the background for the 1-form symmetry $\Gamma^{(1)}$ depends on a choice of background for the 0-form symmetry $\Gamma^{(0)}$. A background for 0-form symmetry is specified by
\be
B_1\in Z^1\left(M_d,\Gamma^{(0)}\right)\,,
\ee
which is a $\Gamma^{(0)}$-valued 1-cocycle on the $d$-dimensional spacetime manifold $M_d$. In the presence of this 0-form background, a background for the 1-form symmetry is specified by
\be
B_2\in Z^2_{B_1}\left(M_d,\Gamma^{(1)}\right)\,,
\ee
which is a $\Gamma^{(1)}$-valued $B_1$-twisted 2-cocycle. See appendix \ref{sec:twistedcoho} for more details.

\subsection{Realisation via Gauge Outer-Automorphisms}
Now let us discuss how outer-automorphisms of gauge algebras provide examples of the situation discussed above, where we have a non-trivial action of the 0-form symmetry on the 1-form symmetry.

\paragraph{Electric 1-form symmetry group.}
Consider a gauge theory with a gauge algebra
\be
\fg=\bigoplus_i\fg_i\,,
\ee
where each $\fg_i$ is a simple, compact, non-abelian Lie algebra. Let the gauge group $\cG$ be a compact, connected Lie group whose Lie algebra is $\fg$. In addition, let us have matter fields $\phi_\alpha$ transforming in irreps $R_\alpha$ of $\cG$. The gauge theory has an electric 1-form symmetry group
\be
\Gamma^{(1)}\subseteq Z_\cG\,,
\ee
which is obtained as the subgroup of the center $Z_\cG$ of $\cG$ that leaves all $R_\alpha$ invariant.

Charged objects are Wilson line defects. A Wilson line in representation $R$ of $\cG$ has charge $q_R\in\wh Z_\cG$ under $Z_\cG$, where $\wh Z_\cG$ is the Pontryagin dual of $Z_\cG$. Its charge under $\Gamma^{(1)}$ is
\be
\wh\pi(q_R)\in\wh\Gamma^{(1)}\,,
\ee
where
\be
\wh\pi:~\wh Z_\cG\to\wh\Gamma^{(1)}\,,
\ee
is the Pontryagin dual of the injective map $\Gamma^{(1)}\to Z_\cG$ making $\Gamma^{(1)}$ a subgroup of $Z_\cG$.

\paragraph{Gauge outer-automorphisms.}
The discussion of 0-form symmetries begins with the group $\cO_\fg$ formed by outer-automorphisms of the gauge algebra $\fg$. This includes not only the outer-automorphisms for each simple factor\footnote{As is well-known and reviewed in appendix \ref{sec:liealgeouterreview}, such outer-automorphisms are classified by permutations of the nodes of the Dynkin diagram of the simple factor $\fg_i$ that leave the Dynkin diagram invariant.} $\fg_i$, but also permutations of simple Lie algebras that are isomorphic to each other. These two types of outer-automorphism do not commute in general, and hence $\cO_\fg$ is a non-abelian finite group in general.

For example, consider
\be
\fg=\so(2m)\oplus\so(2m)\,,
\ee
There is a $\Z_2$ outer-automorphism of each $\so(2m)$ factor. Thus $\Z_2\times\Z_2$ is a subgroup of $\cO_\fg$. We moreover have the exchange of the two $\so(2m)$ factors, which acts on the above subgroup of $\cO_\fg$ by exchanging the two $\Z_2$ subfactors. In total, the outer-automorphism group is
\be
\cO_\fg=\left(\Z_2\times\Z_2\right)\rtimes\Z_2=D_8\,,
\ee
namely the dihedral group of order 8, which is a non-abelian group.

\paragraph{Gauge outer-automorphism symmetries and their action.}
The gauge theory may now realise a subgroup
\be
\Gamma^{(0)}\subseteq\cO_\fg\,,
\ee
of gauge outer-automorphisms as 0-form symmetries. A necessary condition that $\Gamma^{(0)}$ has to satisfy is that it should preserve the Wilson line content of the gauge theory, which means the following. As reviewed in appendix \ref{sec:liealgeouterreview}, there is a natural action of $\cO_\fg$, and hence of $\Gamma^{(0)}$, on representations of $\fg$. Now the above necessary condition states that the action of $\Gamma^{(0)}$ on representations of the gauge algebra $\fg$ should define a closed group action when restricted to representations of the gauge group $\cG$. This is because Wilson lines in a gauge theory with gauge group $\cG$ correspond to representations of $\cG$, and the action of $\Gamma^{(0)}$ should be closed on the Wilson lines.

Computationally, let
\be\label{eq12}
\cG'=\prod_i\cG'_i\,,
\ee
with each $\cG'_i$ being the simply connected, compact Lie group with Lie algebra $\fg_i$. The center $Z_{\cG'}$ of $\cG'$ is
\be
Z_{\cG'}=\prod_i Z_{\cG'_i}\,,
\ee
where $Z_{\cG'_i}$ is the center of $\cG'_i$. Correspondingly, the Pontryagin dual $\wh Z_{\cG'}$ of $Z_{\cG'}$ is
\be
\wh Z_{\cG'}=\prod_i \wh Z_{\cG'_i}\,,
\ee
where $\wh Z_{\cG'_i}$ is the Pontryagin dual of $Z_{\cG'_i}$.
The outer-automorphism group $\cO_\fg$ and hence $\Gamma^{(0)}$ has a well-defined group action on $\wh Z_{\cG'}$ descending from the action on representations of $\fg$. This defines a dual group action of $\Gamma^{(0)}$ on $Z_{\cG'}$. Moreover, we can express
\be\label{eq13}
\cG=\cG'/\cZ_{\cG'}\,,
\ee
where $\cZ_{\cG'}\subseteq Z_{\cG'}$. Then, the above requirement can be restated to say that the action of $\Gamma^{(0)}$ on  $Z_{\cG'}$ restricts to a closed group action on the subgroup $\cZ_{\cG'}\subseteq Z_{\cG'}$.

Thus, if this condition is satisfied, then we have a well-defined group action of $\Gamma^{(0)}$ on $Z_\cG$, because it can be expressed as
\be
Z_\cG=Z_{\cG'}/\cZ_{\cG'}\,,
\ee
and we have closed actions of $\Gamma^{(0)}$ on $Z_{\cG'}$ and $\cZ_{\cG'}$. Dually, we obtain an action of $\Gamma^{(0)}$ on $\wh Z_\cG$.

An example where the above condition is not satisfied is provided by a gauge theory with $Ss(4m)$ gauge group, which is a Lie group with Lie algebra $\so(4m)$, such that the spinor representation of $\so(4m)$ is an allowed representation of $Ss(4m)$, but the cospinor and vector representations of $\so(4m)$ are not allowed representations of $Ss(4m)$. The outer-automorphism group is $\cO_\fg=\Z_2$ (for $m>2$), which exchanges spinor and cospinor representations of $\so(4m)$, and hence does not define a closed group action on representations of $Ss(4m)$. Consequently, such a gauge theory cannot have any 0-form symmetry descending from gauge outer-automorphisms.

Another necessary condition is that the action of $\Gamma^{(0)}$ should preserve the matter content, which means that the action of $\Gamma^{(0)}$ on representations of $\cG$ should descend to a well-defined group action on the set of irreps $R_\alpha$ formed by matter fields.

As a consequence of the above condition being satisfied, we obtain a well-defined group action of $\Gamma^{(0)}$ on the 1-form symmetry group $\Gamma^{(1)}$. This action can be obtained as follows. Each representation $R_\alpha$ has a certain charge $q_\alpha\in\wh Z_\cG$. Collecting all such $q_\alpha$, we obtain a subset $M_\cG$ of $\wh Z_\cG$. Let
\be
\cM_\cG=\text{Span}\left(M_\cG\right)\subseteq\wh Z_\cG\,,
\ee
be the subgroup of $\wh Z_\cG$ spanned by the elements of the subset $M_\cG$. Then, we can identify
\be
\wh\Gamma^{(1)}=\wh Z_\cG/\cM_\cG\,.
\ee
The above condition on the action of $\Gamma^{(0)}$ guarantees that we have a closed group action on $\cM_\cG$, and hence we have a well-defined action on the quotient group $\wh\Gamma^{(1)}$, which can be dualized to provide an action of $\Gamma^{(0)}$ on $\Gamma^{(1)}$.

\paragraph{Example: Pure gauge theory.}
A simple example is provided by a pure non-supersymmetric gauge theory with an arbitrary gauge algebra $\fg$ of the above type, and with gauge group
\be
\cG=\prod_i\cG_i\,,
\ee
where each $\cG_i$ is the simply connected, compact Lie group with Lie algebra $\fg_i$. We have
\be
\Gamma^{(1)}=Z_\cG=\prod_iZ_{\cG_i}\,,
\ee
and
\be
\Gamma^{(0)}=\cO_\fg\,.
\ee
The action of $\Gamma^{(0)}$ on $\Gamma^{(1)}$ can be simply obtained from the natural action of $\Gamma^{(0)}$ on $\wh\Gamma^{(1)}=\wh Z_\cG$.

\paragraph{Example: Mixed gauge outer-automorphisms.}
An interesting situation worth pointing out is when outer-automorphisms of simple gauge algebra factors are not 0-form symmetries individually, but their combination is a 0-form symmetry.

For example, consider an $SU(n)\times SU(2n)$ gauge theory with a complex scalar field $\phi$ in bifundamental representation $\F\otimes\bar\F$ of $SU(n)\times SU(2n)$. Let us assume that the bare Lagrangian has no other terms except the kinetic terms and minimal coupling of $\phi$ to the gauge fields.

We have
\be
\Gamma^{(1)}=\Z_n\,, \quad 
\cO_\fg=\Z_2\times\Z_2\,.
\ee
The first $\Z_2$ factor acts as\,,
\be
\F\otimes\bar\F\to\bar\F\otimes\bar\F
\ee
and hence cannot be promoted to a 0-form symmetry. Similarly, the second $\Z_2$ factor acts as
\be
\F\otimes\bar\F\to\F\otimes\F\,,
\ee
and hence can also not be promoted to a 0-form symmetry. However, the diagonal $\Z_2$ acts as
\be
\F\otimes\bar\F\to\bar\F\otimes\F\,,
\ee
which can be promoted to a 0-form symmetry acting on $\phi$ by complex conjugation. Thus,
\be
\Gamma^{(0)}=\Z_2\,,
\ee
which acts as the combined outer-automorphism of $SU(n)$ and $SU(2n)$ gauge groups. $\Gamma^{(0)}$ acts on $\Gamma^{(1)}$ by sending each element to its inverse.

\paragraph{Gauge bundles in the presence of 1-form background.}
We now want to describe the gauge bundles that a gauge theory sums over when backgrounds for electric 1-form symmetry $\Gamma^{(1)}$ and outer-automorphism 0-form symmetry $\Gamma^{(0)}$ are turned on.

Let us begin with some special cases. Suppose first that the background field $B_1$ for $\Gamma^{(0)}$ is trivial. Then the background field $B_2$ for $\Gamma^{(1)}$ is an element of $Z^2\left(M_d,\Gamma^{(1)}\right)$ without any twist. The gauge theory sums over bundles
\be
A_G:~M_d\to BG\,,
\ee
for the group
\be
G:=\cG/\Gamma^{(1)}\,,
\ee
such that
\be
A_G^*w_2=B_2\,,
\ee
where $A_G^*w_2$ is the pull-back to $M_d$ of a fixed representative
\be
w_2\in Z^2\left(BG,\Gamma^{(1)}\right)\,,
\ee
of the class $[w_2]\in H^2\left(BG,\Gamma^{(1)}\right)$ capturing the obstruction for lifting $G$ bundles to $\cG$ bundles.

\paragraph{Gauge bundles in the presence of 0-form background.}
Now suppose that the background field $B_1\in Z^1\left(M_d,\Gamma^{(0)}\right)$ for $\Gamma^{(0)}$ is non-trivial, but the background field $B_2\in Z^2_{B_1}\left(M_d,\Gamma^{(1)}\right)$ for $\Gamma^{(1)}$ is trivial. The gauge theory sums over bundles
\be
A_{\wt\cG}:~M_d\to B\wt\cG\,,
\ee
for the disconnected Lie group
\be\label{wtcG}
\wt\cG:=\cG\rtimes\Gamma^{(0)}\,,
\ee
obtained by taking semi-direct product\footnote{See appendix \ref{sec:semidirprod} for a review of semi-direct product groups.} of $\cG$ and $\Gamma^{(0)}$, such that
\be
A_{\wt\cG}^*w_1=B_1\,,
\ee
where $A_{\wt\cG}^*w_1$ is the pull-back to $M_d$ of a fixed representative
\be
w_1\in Z^1\left(B\wt\cG,\Gamma^{(0)}\right)\,,
\ee
of the class $[w_1]\in H^1\left(B\wt\cG,\Gamma^{(0)}\right)$ capturing the obstruction to restricting $\wt\cG$ bundles to $\cG$ bundles.

Let us describe the construction of $\wt\cG$ in more detail. We begin with the Lie group $\cG'$ defined in (\ref{eq12}). Notice that we can easily define
\be
\wt\cG'=\cG'\rtimes\Gamma^{(0)}\,,
\ee
since we obtain an action of $\Gamma^{(0)}$ on $\cG'$ from the action of $\Gamma^{(0)}\subseteq\cO_\fg$ on $\fg$ by using the exponential map from $\fg$ to $\cG'$. Now, since, as we discussed earlier, $\Gamma^{(0)}$ has a closed action on $\cZ_{\cG'}$, we can use (\ref{eq13}) to argue that the action of $\Gamma^{(0)}$ on $\cG'$ projects to a well-defined action of $\Gamma^{(0)}$ on $\cG$, using which we define $\wt\cG$ via (\ref{wtcG}). Note that $\cZ_{\cG'}$ is a normal subgroup of $\wt\cG'$, because $\cZ_{\cG'}$ is closed under the action of $\Gamma^{(0)}$. We can thus express $\wt\cG$ as the quotient of $\wt\cG'$ by $\cZ_{\cG'}$
\be
\wt\cG=\wt\cG'/\cZ_{\cG'}\,.
\ee

\paragraph{Gauge bundles in the presence of both 0-form and 1-form backgrounds.}
Let us now turn on a background field $B_1\in Z^1\left(M_d,\Gamma^{(0)}\right)$ for $\Gamma^{(0)}$, and a background field $B_2\in Z^2_{B_1}\left(M_d,\Gamma^{(1)}\right)$ for $\Gamma^{(1)}$. The gauge theory sums over bundles
\be
A_{\wt G}:~M_d\to B\wt G\,,
\ee
for the disconnected Lie group
\be
\wt G:=G\rtimes\Gamma^{(0)}=\wt\cG/\Gamma^{(1)}\,,
\ee
obtained by modding out the normal subgroup $\Gamma^{(1)}$ of $\wt\cG$, such that
\be
A_{\wt G}^*w_1=B_1\,,
\ee
where $A_{\wt G}^*w_1$ is the pull-back to $M_d$ of a fixed representative
\be
w_1\in Z^1\left(B\wt G,\Gamma^{(0)}\right)\,,
\ee
of the class $[w_1]\in H^1\left(B\wt G,\Gamma^{(0)}\right)$ capturing the obstruction to restricting $\wt G$ bundles to $G$ bundles, and
\be
A_{\wt G}^*w_2=B_2\,,
\ee
where $A_{\wt G}^*w_2$ is the pull-back to $M_d$ a fixed representative
\be
w_2\in Z^2_{w_1}\left(B\wt G,\Gamma^{(1)}\right)\,,
\ee
of the class $[w_2]\in H^2_{w_1}\left(B\wt G,\Gamma^{(1)}\right)$ capturing the obstruction for lifting $\wt G$ bundles to $\wt\cG$ bundles.

\paragraph{Subtleties.}
The 0-form symmetry $\Gamma^{(0)}$ is actually a subgroup of automorphisms $\Aut(\fg)$ of $\fg$ which projects to the $\Gamma^{(0)}$ subgroup of outer-automorphisms $\cO_\fg$ of $\fg$ that we have been discussing above. The specification of $\Gamma^{(0)}$ as a subgroup of $\cO_\fg$ only specifies its action on the Wilson line defects, but to specify the action of $\Gamma^{(0)}$ on (genuine and non-genuine) local operators, we have to specify $\Gamma^{(0)}$ as a subgroup of $\Aut(\fg)$.

One might expect that different choices of $\Gamma^{(0)}\subseteq \Aut(\fg)$ that project to the same $\Gamma^{(0)}\subseteq \cO_\fg$ will lead to physically equivalent results, as such different choices $\Gamma^{(0)}\subseteq \Aut(\fg)$ are related by inner automorphisms of $\fg$. However, it is known that this expectation is not correct \cite{Arias-Tamargo:2019jyh,Arias-Tamargo:2021ppf,Henning:2021ctv}.

For example, the choice of $\Gamma^{(0)}\subseteq \Aut(\fg)$ enters into the semi-direct products defining disconnected Lie groups $\wt\cG'$, $\wt\cG$ and $\wt G$ discussed above. Now, different choices $\Gamma^{(0)}_a\simeq\Gamma^{(0)}\subseteq \Aut(\fg)$ that project to the same $\Gamma^{(0)}\subseteq \cO_\fg$ lead to disconnected groups $\wt\cG'_a=\cG'\rtimes\Gamma^{(0)}_a$ such that
\be
\wt\cG'_a\not\simeq\wt\cG'_b\,,
\ee
in general. And similar statements hold in general for $\wt\cG_a$ and $\wt G_a$.

A well-known example is provided by
\be
\fg=\su(2n);\quad\cG'=SU(2n)\,,
\ee
and
\be
\Gamma^{(0)}=\cO_\fg=\Z_2\,,
\ee
for which there are two possible disconnected groups
\be
\wt\cG'=\wt{SU}(2n)_I,~\wt{SU}(2n)_{II}\,,
\ee
as explained in \cite{Arias-Tamargo:2019jyh,Arias-Tamargo:2021ppf}.

The choice of $\Gamma^{(0)}_a\subseteq \Aut(\fg)$ reflects itself in the action of $\Gamma^{(0)}$ on (genuine and non-genuine) local operators, which form representations of the corresponding disconnected group $\wt\cG_a$. Moreover, in even spacetime dimension, and in the presence of fermions, some choices of $\Gamma^{(0)}_a\subseteq \Aut(\fg)$ may suffer from ABJ anomalies and hence be disallowed quantum mechanically, while some other choices of $\Gamma^{(0)}_a\subseteq \Aut(\fg)$ may not suffer from ABJ anomalies and hence allow $\Gamma^{(0)}$ to be a 0-form symmetry of the theory quantum mechanically. See \cite{Henning:2021ctv} for more details.

In this paper, we ignore such subtleties and work at the level of outer-automorphisms, leaving the proper accounting of such subtleties to future work.

\subsection{Incorporating 0-Form Outer-Automorphisms}\label{0fa1f0f}
In the previous subsection, we saw that outer-automorphisms of gauge algebras provide natural examples of 0-form symmetries acting on 1-form symmetries, because they act on Wilson line defects which are objects charged under electric 1-form symmetries. In this subsection, we consider outer-automorphisms of Lie algebras associated to continuous 0-form symmetries. Such outer-automorphisms may need to be combined with gauge outer-automorphisms for them to descend to 0-form symmetries of the theory. In such a situation, the resulting 0-form symmetries can act on 1-form symmetries, as their definition involves gauge outer-automorphisms, which as discussed above provide a natural action.

\paragraph{General outer-automorphisms.}
Let us consider a gauge theory as in the previous subsection. Let its continuous flavor 0-form symmetries be described by a Lie algebra
\be\label{contf}
\ff=\bigoplus_a\ff_a\,,
\ee
where each $\ff_a$ is either a simple, compact, non-abelian Lie algebra, or a $\u(1)$ factor. As for the gauge algebra $\fg$, we define $\cO_\ff$ to be the group of outer-automorphisms of $\ff$. Thus, the total group of outer-automorphisms is
\be
\cO=\cO_\fg\times\cO_\ff\,.
\ee

\paragraph{General outer-automorphism symmetries.}
A subgroup
\be
\Gamma^{(0)}\subseteq\cO\,,
\ee
may be realised as 0-form symmetry of the theory. The 0-form symmetry group $\Gamma^{(0)}_\fg$ comprising of gauge outer-automorphisms studied in the previous subsection (where it was denoted simply as $\Gamma^{(0)}$) can be obtained as
\be
\Gamma^{(0)}_\fg=\pi_\fg\left(\Gamma^{(0)}\right)\,,
\ee
where
\be
\pi_\fg:~\cO\to\cO_\fg\,,
\ee
is the projection map obtained by forgetting $\cO_\ff$. 

\paragraph{Example: Mixing of gauge and 0-form outer-automorphisms.}
In particular, some of the gauge outer-automorphisms may need to be supplemented by 0-form outer-automorphisms to form symmetries. For example, consider a gauge theory with $SU(n)$ gauge group and $m$ complex scalar fields, each transforming in fundamental representation of $SU(n)$. Then, we clearly have
\be\label{eq4}
\Gamma^{(0)}_\fg=\cO_\fg=\Z_2\,.
\ee
However, let us now take into account the information about the 0-form symmetry algebra $\su(m)\oplus\u(1)$ rotating the $m$ complex scalars. Now, we would say that we have a complex scalar transforming in bifundamental representation of $\su(n)\oplus\su(m)$ and has charge 1 under $\u(1)$, where $\su(n)$ is the gauge algebra and $\su(m),\u(1)$ are 0-form symmetry algebras. We have
\be
\cO=\cO_\fg\times\cO_\ff=\cO_{\su(n)}\times\cO_{\su(m)}\times\cO_{\u(1)}=\Z_2\times\Z_2\times\Z_2\,,
\ee
but only the diagonal of the three $\Z_2$ leaves the matter content invariant. That is, the gauge outer-automorphism of $\su(n)$ needs to be supplemented by the 0-form outer-automorphisms of $\su(m)$ and $\u(1)$ for it to be a 0-form symmetry of the theory. Thus, we have
\be
\Gamma^{(0)}=\Z_2\subset\cO\,,
\ee
and indeed
\be
\Gamma^{(0)}_\fg=\pi_\fg\left(\Gamma^{(0)}\right)\,.
\ee

\paragraph{Purely 0-form outer-automorphism symmetries.}
Let us promote the projection map $\Gamma^{(0)}\to \Gamma^{(0)}_\fg$ to a short exact sequence
\be
1\to\Gamma^{(0)}_\ff\to\Gamma^{(0)}\to\Gamma^{(0)}_\fg\to1 \,,
\ee
where $\Gamma^{(0)}_\ff\subseteq\Gamma^{(0)}$ is the kernel of the projection map $\Gamma^{(0)}\to \Gamma^{(0)}_\fg$, and can also be described as
\be
\Gamma^{(0)}_\ff=\Gamma^{(0)}\cap\cO_\ff\,.
\ee
That is, $\Gamma^{(0)}_\ff$ is the subgroup of outer-automorphism symmetry group $\Gamma^{(0)}$ whose elements comprise purely of 0-form outer-automorphisms without any additional supplementation with gauge outer-automorphisms.

\paragraph{Example: Non-trivial $\Gamma^{(0)}_\ff$.}
Consider a gauge theory with gauge group $\cG=\Sp(n)$ and 0-form flavor symmetry algebra $\ff=\so(2m)$ such that we have a scalar field transforming in fundamental representation of $\Sp(n)$ and vector representation of $\so(2m)$. Then, we have
\be\label{eq3}
\Gamma^{(0)}=\Gamma^{(0)}_\ff=\cO_\ff=\Z_2\,.
\ee
That is, we have an outer-automorphism symmetry descending from the non-trivial outer-automorphism of $\ff=\so(2m)$, which does not need to be supplemented by a gauge outer-automorphism.

\paragraph{Action on electric 1-form symmetry group.}
The outer-automorphism 0-form symmetry group $\Gamma^{(0)}$ acts on the electric 1-form symmetry group $\Gamma^{(1)}$ only via its projection $\Gamma^{(0)}_\fg$, which as discussed in the previous subsection has a natural action on $\Gamma^{(1)}$. In particular, the purely 0-form outer-automorphism symmetry group $\Gamma^{(0)}_\ff$ does not act on $\Gamma^{(1)}$.

\section{Disconnected 0-Form Symmetry Groups}
\label{sec:disconnected0fs}
In this section, we continue the study of outer-automorphism symmetries arising from outer-automorphisms of 0-form symmetry Lie algebras. We discuss how such symmetries promote the 0-form symmetry group to a disconnected Lie group.

\subsection{Review: Connected Part of 0-Form Symmetry Group}
\paragraph{Definition.}
Consider a general theory $\fT$ in any spacetime dimension $d$. It may have a continuous 0-form symmetry associated to a Lie algebra $\ff$ as discussed around (\ref{contf}). Let us assume in what follows that there exists a genuine local operator transforming in a non-trivial representation of each $\ff_a$ factor. This is an assumption regarding the $\u(1)$ factors only, because for a non-abelian factor the current operator $j_{\mu,a}$ associated to $\ff_a$ provides such an operator, as it transforms in the adjoint representation of $\ff_a$ which is a non-trivial representation if $\ff_a$ is non-abelian.

Global information regarding such a continuous 0-form symmetry is captured by the specification of a Lie group $\cF$, which we refer to as the \textit{connected part of 0-form symmetry group}. The group $\cF$ satisfies the following properties:
\ben
\item $\cF$ is a compact, connected Lie group with Lie algebra $\ff$.
\item All genuine local operators transform in representations of $\ff$ that are also representations of $\cF$.
\item There exists at least one genuine local operator transforming in each representation of $\cF$.
\een

\paragraph{Example.}
Consider a $U(1)$ gauge theory with two complex scalar fields $\phi_1$ and $\phi_2$, each having same charge $q$ under the $U(1)$ gauge group. Let the theory have a continuous 0-form symmetry 
\be
\ff=\su(2)\,,
\ee
under which $\phi_i$ form a doublet. The genuine local operators comprise of gauge invariant combinations of $\phi_1$ and $\phi_2$, which only give rise to integer spin representations of $\ff=\su(2)$. Consequently, the connected part of 0-form symmetry group is
\be
\cF=SO(3)\,.
\ee

\paragraph{Backgrounds.}
A background for a continuous 0-form symmetry is specified by:
\ben
\item A map
\be
A_\cF:~M_d\to B\cF\,,
\ee
using which we can pull-back the principal $\cF$-bundle $E\cF$ on $B\cF$ to a principal $\cF$-bundle $A_\cF^*(E\cF)$ on spacetime $M_d$.
\item A connection on $A_\cF^*(E\cF)$.
\een
Thus, the 0-form symmetry group $\cF$ informs the possible background bundles for the continuous 0-form symmetry.

In the above example, this means that we can turn on background $SO(3)$ bundles, which are more general than $SU(2)$ bundles because $SU(2)$ bundles can be recognized as $SO(3)$ bundles with $[w_2]=0$, where $[w_2]$ is the second Stiefel-Whitney class associated to an $SO(3)$ bundle.

\paragraph{Including non-genuine local operators.}
As we have discussed above, all genuine local operators form representations of $\cF$. However, non-genuine local operators may form representations of $\ff$ that are not representations of $\cF$, but instead are projective representations of $\cF$. Let us introduce an auxiliary group $F$ with the following properties:
\ben
\item $F$ is a compact, connected Lie group with Lie algebra $\ff$.
\item All genuine and non-genuine local operators form representations of $\ff$ that are also representations of $F$.
\item There exists at least one genuine or non-genuine local operator transforming in each representation of $F$.
\een
We can relate $F$ to the 0-form symmetry group $\cF$ via
\be
\cF=F/\cZ\,,
\ee
where $\cZ$ is a subgroup of the center $Z_F$ of $F$.

In the above example, the matter fields $\phi_i$ provide examples of gauge invariant non-genuine local operators that do not transform in representations of $SO(3)$, since they form a doublet of $\su(2)$. See figure \ref{ngngo}. Thus we find that $F=SU(2)$ and $\cZ=Z_F=Z_{SU(2)}=\Z_2$.

\begin{figure}
\centering
\begin{tikzpicture}
\draw[thick] (-4.5,-0.5) -- (0,-0.5);
\node[] at (-5,-0.5) {$W_q$};
\node[red] at (0.5,-0.5) {$\phi_i$};
\draw [red,fill=red] (0,-0.5) ellipse (0.05 and 0.05);
\end{tikzpicture}
\caption{The local operators corresponding to matter fields $\phi_i$ are non-gauge invariant local operators, but can be made gauge invariant by inserting them at the end of a Wilson line defect $W_q$ of charge $q$ under the $U(1)$ gauge group. Thus $\phi_i$ give rise to well-defined gauge-invariant non-genuine local operators.}
\label{ngngo}
\end{figure}

\paragraph{Background field valued in $\cZ$.}
To each background bundle $A_\cF$, we can associate a background field
\be
A_\cF^*w_2\in Z^2(M_d,\cZ)\,,
\ee
which is a 2-cocycle on spacetime obtained as the pull-back of a fixed representative
\be
w_2\in Z^2(B\cF,\cZ)\,,
\ee
of the characteristic class $[w_2]\in H^2(B\cF,\cZ)$ capturing the obstruction of lifting $\cF$ bundles to $F$ bundles.

In the above example, $[w_2]$ is the second Stiefel-Whitney class associated to $SO(3)$ bundles.

\paragraph{Background Wilson lines.}
Consider the insertion of a genuine local operator $O$ in the presence of a non-trivial background for $\cF$. Suppose $O$ transforms in a representation $R$ of $\cF$. This means that the insertion of $O$ is not gauge invariant under background gauge transformations for $\cF$. To obtain a well-defined gauge invariant insertion, we need to attach $O$ to a background Wilson line in representation $R$ of $\cF$. Similarly, to a non-genuine local operator transforming in a representation $R$ of $F$, we need to attach a background Wilson line in representation $R$ of $F$. 

To a background Wilson line in an irreducible representation $R$ of $F$, we can associate an element
\be\label{eq7}
q_R\in\wh\cZ
\ee
describing the charge of the representation $R$ under the subgroup $\cZ$ of the center $Z_F$ of $F$. Note that, if $R$ is also a representation of $\cF$, rather than being a projective representation, then it has $q_R=0$.

\paragraph{Construction of $F$ and $\cF$.}
We can construct $F$ starting from a group $F'$ defined as
\be\label{eq6}
F':=\prod_a F'_a\,,
\ee
where $F'_a$ is the compact, simply connected group with Lie algebra $\ff_a$, if $\ff_a$ is non-abelian; and if $\ff_a=\u(1)$, then $F'_a\simeq U(1)$ is a group with periodicity chosen to be such that all genuine and non-genuine local operators have non-fractional integer charges under $F'_a$. Let $Z_{F'}$ be the center of $F'$, which can be expressed as
\be
Z_{F'}=\prod_a Z_{F'_a}\,,
\ee
where $Z_{F'_a}$ is the center of $F'_a$. Similarly, the Pontryagin dual $\wh Z_{F'}$ of $Z_{F'}$ can be expressed as
\be
\wh Z_{F'}=\prod_a \wh Z_{F'_a}\,,
\ee
where $\wh Z_{F'_a}$ is the Pontryagin dual of $Z_{F'_a}$.

We can compute $F$ in terms of $F'$ as
\be
F=F'/\cZ_{F'}\,,
\ee
where $\cZ_{F'}$ is a subgroup of $Z_{F'}$ obtained as the Pontryagin dual of
\be
\wh\cZ_{F'}:=\wh Z_{F'}/\cM_{F'}\,,
\ee
where
\be
\cM_{F'}=\text{Span}(M_{F'})\subseteq\wh Z_{F'}\,,
\ee
is obtained by taking the span in $\wh Z_{F'}$ of the elements lying in the subset $M_{F'}$ of $\wh Z_{F'}$ obtained by collecting the charges under $Z_{F'}$ of all the genuine and non-genuine local operators. The center $Z_F$ of $F$ is
\be
Z_F=Z_{F'}/\cZ_{F'}\,,
\ee
which is the Pontryagin dual of $\cM_{F'}$. Similarly, we can compute $\cF$ in terms of $F$ as
\be
\cF=F/\cZ\,,
\ee
where $\cZ$ is a subgroup of $Z_{F}$ obtained as the Pontryagin dual of
\be
\wh\cZ:=\wh Z_{F}/\cM_{F}\,,
\ee
where
\be
\cM_{F}=\text{Span}(M_{F})\subseteq\wh Z_{F}\,,
\ee
is obtained by taking the span in $\wh Z_{F}$ of the elements lying in the subset $M_{F}$ of $\wh Z_{F}$ obtained by collecting the charges under $Z_{F}$ of all the genuine local operators. 

\subsection{Finite 0-Form Symmetries and Outer-Automorphisms}
\paragraph{Relationship between finite and continuous 0-form symmetries.}
In addition to continuous 0-form symmetries, we can also have a finite 0-form symmetry group $\Gamma^{(0)}$ which may be non-abelian. In general, it is possible for some elements of $\Gamma^{(0)}$ to act on all genuine and non-genuine local operators in exactly the same way as some elements of $F$. Let $\Gamma$ be the group formed by such elements. Then, the combined group of symmetries acting on genuine and non-genuine local operators is
\be
\frac{F\rtimes\Gamma^{(0)}}{\Gamma}\,,
\ee
where the semi-direct product appears because some elements of $\Gamma^{(0)}$ might act on $F$ via outer-automorphisms.

In this paper, we focus on the situation where $\Gamma$ is trivial, so that the combined group of symmetries acting on genuine and non-genuine local operators is
\be
\wt F:=F\rtimes\Gamma^{(0)}\,.
\ee
Let us describe the construction of this group $\wt F$.

\paragraph{Outer-automorphisms associated to finite 0-form symmetries.}
First of all, given an element $o\in\Gamma^{(0)}$, we can associate to it an element of the group $\cO_\ff$ of outer-automorphisms of 0-form symmetry algebra $\ff$, using the action of $o$ on the current operator $j^\mu$ of $\ff$, as this action captures the action of $o$ on the Lie algebra $\ff$ itself. This map provides a homomorphism
\be\label{piff}
\pi_\ff:~\Gamma^{(0)}\to\cO_\ff\,,
\ee
describing the 0-form outer-automorphism associated to each element in $\Gamma^{(0)}$.

\paragraph{Construction of $\wt F$.}
This provides us with an action of $\Gamma^{(0)}$ on the Lie group $F'$ defined in (\ref{eq6}), which is obtained by extending the action of $\Gamma^{(0)}$ on $\ff$ using the exponential map. This allows us to define a disconnected Lie group
\be
\wt F'=F'\rtimes\Gamma^{(0)}\,,
\ee
whose connected part is $F'$.

Now, we claim that this action of $\Gamma^{(0)}$ on $F'$ descends to well-defined action of $\Gamma^{(0)}$ on $F$. To show this, it is sufficient to show that the action of $\Gamma^{(0)}$ on $F'$ restricts to a closed action on the subgroup $\cZ_{F'}$ of $F'$. This essentially follows from the fact that $\Gamma^{(0)}$ must take local operators into local operators. This requirement leads to a closed action on the set $M_{F'}$, which extends to a closed action on the group $\cM_{F'}$. Now, notice that from the action of $\Gamma^{(0)}$ on $F'$, we obtain a closed action of $\Gamma^{(0)}$ on the subgroup $Z_{F'}$ of $F'$, which can be dualized to an action on $\wh Z_{F'}$. The fact that we have a closed action on $\cM_{F'}\subseteq\wh Z_{F'}$ means that the action on $\wh Z_{F'}$ projects to a well-defined action of $\Gamma^{(0)}$ on $\wh\cZ_{F'}$, leading to a dual action on $\cZ_{F'}$ as required. 

Thus we have defined $\wt F$, which is a disconnected Lie group that can be written in multiple ways as follows
\be
\wt F=F\rtimes\Gamma^{(0)}=\wt F'/\cZ_{F'}\,,
\ee
and whose connected part is the group $F$. Note that $\wt F'/\cZ_{F'}$ is a well-defined quotient because $\cZ_{F'}$ is a normal subgroup of $\wt F'$, which follows from the fact that $\Gamma^{(0)}$ has a closed action on $\cZ_{F'}$ as discussed above.

\subsection{Disconnected 0-Form Symmetry Group}
\paragraph{Construction.}
In a similar fashion as above, we can show that the action of $\Gamma^{(0)}$ on $F$ descends to a well-defined action on $\cF$, where the essential part of the argument is that $\Gamma^{(0)}$ must take genuine local operators into genuine local operators. In more detail, the above requirement leads to a closed action on the set $M_{F}$, which extends to a closed action on the group $\cM_{F}$. Now, notice that from the action of $\Gamma^{(0)}$ on $F$, we obtain a closed action of $\Gamma^{(0)}$ on the subgroup $Z_{F}$ of $F$, which can be dualized to an action on $\wh Z_{F}$. The fact that we have a closed action on $\cM_{F}\subseteq\wh Z_{F}$ means that the action on $\wh Z_{F}$ projects to a well-defined action of $\Gamma^{(0)}$ on $\wh\cZ$, leading to a dual action on $\cZ\subseteq F$, which implies that the action of $\Gamma^{(0)}$ on $F$ projects to a well-defined action on $\cF$.

We can thus define a disconnected Lie group that can be written in multiple ways as follows
\be
\wt\cF:=\cF\rtimes\Gamma^{(0)}=\wt F/\cZ\,,
\ee
and whose connected part is the group $\cF$. Note that $\wt F/\cZ$ is a well-defined quotient because $\cZ$ is a normal subgroup of $\wt F$, which follows from the fact that $\Gamma^{(0)}$ has a closed action on $\cZ$ as discussed above.

\paragraph{Interpretation.}
We have seen in this section that the backgrounds for continuous 0-form symmetry are specified by the group $\cF$, while the backgrounds for finite 0-form symmetry have to be specified by the group $\Gamma^{(0)}$. Thus, the full 0-form symmetry backgrounds are specified by the above disconnected Lie group $\wt\cF$, which we call as the full \textit{0-form symmetry group} of the theory.

\paragraph{Backgrounds.}
A background for full 0-form symmetry, including both continuous and finite parts, is specified by:
\ben
\item A map
\be
A_{\wt\cF}:~M_d\to B\wt\cF\,,
\ee
using which we can pull-back the principal $\wt\cF$-bundle $E\wt\cF$ on $B\wt\cF$ to a principal $\wt\cF$-bundle $A_{\wt\cF}^*(E\wt\cF)$ on spacetime $M_d$.
\item A connection on $A_{\wt\cF}^*(E\wt\cF)$.
\een

\ni Such a bundle comes equipped with two special background fields:
\ben
\item First, we have a $\Gamma^{(0)}$-valued background field
\be
A_{\wt\cF}^*w_1\in Z^1\left(M_d,\Gamma^{(0)}\right)\,,
\ee
which is a 1-cocycle on spacetime obtained as the pull-back of a fixed representative
\be
w_1\in Z^1\left(B\wt\cF,\Gamma^{(0)}\right)\,,
\ee
of the characteristic class $[w_1]\in H^1\left(B\wt\cF,\Gamma^{(0)}\right)$ capturing the obstruction for restricting $\wt\cF$ bundles to $\cF$ bundles. This field is recognized as the background field $B_1$ for the finite 0-form symmetry group $\Gamma^{(0)}$
\be
A_{\wt\cF}^*w_1=B_1\,.
\ee
\item Second, we have a $\cZ$-valued background field
\be
A_{\wt\cF}^*w_2\in Z^2_{B_1}\left(M_d,\cZ\right)\,,
\ee
which is a $B_1$-twisted 2-cocycle on spacetime obtained as the pull-back of a fixed representative
\be
w_2\in Z^2_{w_1}\left(B\wt\cF,\cZ\right)\,,
\ee
of the characteristic class $[w_2]\in H^2_{w_1}\left(B\wt\cF,\cZ\right)$ capturing the obstruction for lifting $\wt\cF$ bundles to $\wt F$ bundles.
\een

\subsection{0-Form Symmetry Groups in Gauge Theories}
\label{sec:disconflav}
Let us return to the discussion of subsection \ref{0fa1f0f} regarding gauge theories and describe the computation of the disconnected 0-form symmetry group $\wt\cF$ for such theories, where the continuous 0-form symmetry being considered is a flavor symmetry with flavor algebra $\ff$.

\paragraph{Determining the connected part $\cF$ of 0-form symmetry group.}
Our first task is to compute the connected part $\cF$ of $\wt\cF$, for which we need to first compute $F$ from $F'$. As discussed above, the only input for this computation is the subset $M_{F'}$ of $\wh Z_{F'}$, which in our gauge-theoretic context, is provided by the charges under $Z_{F'}$ of all matter fields.

In order to compute $\cF$, we have to compute the subset $M_F$ of charges in $\wh Z_F$ occupied by genuine local operators, which are gauge invariant combinations of matter fields. For this purpose, we collect the charges of matter fields under $Z_\cG\times Z_F$, which form a subset $M$ of $\wh Z_\cG\times\wh Z_F$. Then, we have
\be\label{eq11}
M_F=M\cap\wh Z_F\,.
\ee

\paragraph{Determining the full 0-form symmetry group $\wt\cF$.}
Now, we include the finite 0-form symmetry group $\Gamma^{(0)}\subseteq\cO_\fg\times\cO_\ff$ discussed in subsection \ref{0fa1f0f}. Then, the homomorphism (\ref{piff}) is provided simply by restricting the projection map $\cO_\fg\times\cO_\ff\to\cO_\ff$, which forgets the information about $\cO_\fg$, to the subgroup $\Gamma^{(0)}\subseteq\cO_\fg\times\cO_\ff$. 

As discussed above, using the data of this homomorphism, we obtain the disconnected Lie group $\wt F$ and the 0-form symmetry group $\wt\cF$.

\paragraph{Example 1.}
Consider the gauge theory discussed around (\ref{eq3}). For this theory, we have
\be
\ff=\so(2m)\,,
\ee
and so
\be
F'=\Spin(2m)\,.
\ee
Since we have matter fields transforming in vector representation of $\so(2m)$, we find that
\be
F=SO(2m)\,.
\ee
The matter field has non-trivial charge under
\be
Z_F=\Z_2\,,
\ee
but also a non-trivial charge under
\be
Z_\cG=Z_{\Sp(n)}=\Z_2\,.
\ee
Thus, $\wh Z_F\cap M$ is trivial, i.e. the gauge invariant operators have trivial charge under $Z_F$. Hence, we have
\be
\wh\cZ=\Z_2\,,
\ee
implying that
\be
\cZ=\Z_2=Z_F\,,
\ee
and so the connected part of 0-form symmetry group associated to the flavor symmetry $\ff=\so(2m)$ is
\be
\cF=SO(2m)/\Z_2=PSO(2m)\,.
\ee
As we discussed around (\ref{eq3}), the non-trivial outer-automorphism of $\so(2m)$ flavor algebra is also a symmetry of the theory. Thus, we have
\be
\wt F = SO(2m)\rtimes\Z_2=O(2m)\,,
\ee
and the full 0-form symmetry group $\wt\cF$ combining the continuous $\so(2m)$ symmetry and the $\Z_2$ outer-automorphism symmetry is given by
\be
\wt\cF=PSO(2m)\rtimes\Z_2=PO(2m)\,.
\ee
The first Stiefel-Whitney class $[w_1]$ of $PO(2m)$ bundles describes the obstruction to restricting $PO(2m)$ bundles to $PSO(2m)$ bundles, and hence provides the background field for the $\Z_2$ outer-automorphism symmetry of the theory.

\paragraph{Example 2.}
Now consider the gauge theory discussed around (\ref{eq4}). We have
\be
\ff=\su(m)\oplus\u(1)\,,
\ee
for which
\be
F'=SU(m)\times U(1)\,,
\ee
where the periodicity of the $U(1)$ component is chosen such that the matter field has charge 1 under it. Since the matter field is also charged as fundamental of $SU(m)$, we have
\be
F=\frac{SU(m)\times U(1)}{\Z_m}=U(m)\,.
\ee
The matter field has charge 1 under the center
\be
Z_F=U(1)\,,
\ee
of $F=U(m)$, but gauge invariant operators constructed out of it can only have charges under $Z_F$ which are multiples of $n$. Thus, the 0-form symmetry group associated to the continuous flavor symmetry is
\be\label{eq5}
\cF=U(m)/\Z_n\,.
\ee
As we discussed around (\ref{eq4}), the gauge theory also has a $\Z_2$ outer-automorphism symmetry which is the diagonal combination of the non-trivial outer-automorphisms of the gauge algebra $\su(n)$ and flavor algebras $\su(m),\u(1)$. Although outer-automorphisms of $\su(m)$ and $\u(1)$ are individually not outer-automorphisms of the group $U(m)$, their diagonal combination is a $\Z_2$ outer-automorphism of $U(m)$. Thus, the above results are consistent so far, and we have
\be
\wt F=U(m)\rtimes\Z_2=\wt U(m)\,.
\ee
This outer-automorphism of $U(m)$ acts on its $U(1)$ center by sending each element to its inverse. Thus, it has a non-trivial, but closed action on the $\Z_n$ group appearing in the denominator of (\ref{eq5}). As a consequence, the outer-automorphism of $U(m)$ descends to a non-trivial outer-automorphism of $\cF=U(m)/\Z_n$. The total 0-form symmetry group combining the continuous flavor symmetry and the outer-automorphism symmetry is the following disconnected Lie group
\be
\wt\cF=\frac{U(m)}{\Z_n}\rtimes\Z_2=\frac{\wt U(m)}{\Z_n}\,,
\ee
where $\wt U(m)=U(m)\rtimes\Z_2$ is a disconnected group obtained by combining $U(m)$ with its non-trivial outer-automorphism, and $\Z_n$ is a normal subgroup of $\wt U(m)$.

\section{Disconnected 2-Group Symmetries}
\label{sec:discon2groups}
So far, we have seen in the previous sections that the finite part $\Gamma^{(0)}$ of the 0-form symmetry group $\wt\cF$ can act both on the 1-form symmetry group $\Gamma^{(1)}$ and the connected part $\cF$ of the 0-form symmetry group. In many theories, including gauge theories of the type discussed above, $\Gamma^{(1)}$ and $\cF$ mix together to form a 2-group $\fG$, and correspondingly we say that the 1-form symmetry and continuous 0-form symmetry combine to form a 2-group symmetry. Combining the action of $\Gamma^{(0)}$ on $\Gamma^{(1)}$ and $\cF$, we should obtain an action of $\Gamma^{(0)}$ on $\fG$. Thus, the full symmetry structure incorporating $\Gamma^{(0)}$ and $\fG$ should be a 2-group $\wt\fG$ constructed as a ``semi-direct product of a 2-group with a finite group''. We refer to such 2-groups as \textit{disconnected 2-groups}, and the corresponding symmetry structure as a \textit{disconnected 2-group symmetry}. 

In this section we discuss special types of disconnected 2-group symmetries and show that they arise in gauge theories in any spacetime dimension.

\subsection{Review: Connected Part of 2-Group Symmetry}
\paragraph{General 2-group symmetries.}
A general 2-group symmetry $\fG$ mixing 1-form and \textit{continuous} 0-form symmetries is a tuple
\be\label{2gG}
\fG=\left(\Gamma^{(1)},\cF,[\Theta]\right)\,,
\ee
where $\Gamma^{(1)}$ is the 1-form symmetry group, $\cF$ is connected part of 0-form symmetry group, and
\be
[\Theta]\in H^3\left(B\cF,\Gamma^{(1)}\right)\,,
\ee
is known as the \textit{Postnikov class} associated to the 2-group symmetry. Physically, the Postnikov class encodes the non-closed of the background field $B_2$ for 1-form symmetry via
\be
\delta B_2+A_\cF^*\Theta=0\,,
\ee
where
\be
A_\cF:~M_d\to B\cF\,,
\ee
describes the background bundle for the 0-form symmetry group $\cF$, and 
\be
\Theta\in Z^3\left(B\cF,\Gamma^{(1)}\right)\,,
\ee
is a fixed $\Gamma^{(1)}$-valued 3-cocycle on $B\cF$ lying in the class $[\Theta]$.

\paragraph{Bockstein 2-group symmetries.}
In this paper, we focus on particular types of 2-group symmetries, which we call ``Bockstein 2-group symmetries'', for which we have the relationship
\be\label{B2g}
\delta B_2+A_\cF^*\text{Bock}(w_2)=0\,,
\ee
where
\be
w_2\in Z^2(B\cF,\cZ)\,,
\ee
is a cocycle lying in the characteristic class
\be
[w_2]\in H^2(B\cF,\cZ)\,,
\ee
capturing the obstruction for lifting $\cF$ bundles to $F$ bundles. That is, for a Bockstein 2-group symmetry, the Postnikov class can be expressed as
\be
[\Theta]=\text{Bock}\big([w_2]\big)\,,
\ee
where
\be
\text{Bock}:~H^2(B\cF,\cZ)\to H^3\left(B\cF,\Gamma^{(1)}\right)\,,
\ee
is the Bockstein homomorphism in the long exact sequence for cohomology associated to a short exact sequence
\be\label{sesN}
0\to\Gamma^{(1)}\to\cE\to\cZ\to0\,.
\ee
If the above short exact sequence splits, then the Bockstein homomorphism is necessarily trivial and consequently we do not have a 2-group symmetry.

\begin{figure}
\centering
\scalebox{1}{\begin{tikzpicture}
\draw [thick](-4,0) -- (-2,0);
\node at (-4.5,0) {$L_1$};
\node at (0.5,0) {$L_2$};
\node[red] at (-2,-0.5) {$O$};
\node at (-5,1) {(1)};
\draw [thick](0,0) -- (-2,0);
\draw [ultra thick,white](-3.4,0) -- (-3.2,0);
\draw [thick,blue] (-3,0) ellipse (0.3 and 0.7);
\draw [ultra thick,white](-2.7,-0.1) -- (-2.7,0.1);
\draw [red,fill=red] (-2,0) node (v1) {} ellipse (0.05 and 0.05);
\draw [dashed,thick](-2.7396,0) -- (-2.6396,0);
\node[blue] at (-3,-1) {$\gamma\in\Gamma^{(1)}$};
\begin{scope}[shift={(8.8,0)}]
\draw [thick](-4,0) -- (-2,0);
\node at (-4.5,0) {$L_1$};
\draw [thick](0,0) -- (-2,0);
\draw [red,fill=red] (-2,0) node (v1) {} ellipse (0.05 and 0.05);
\node at (0.5,0) {$L_2$};
\node[red] at (-2,-0.5) {$O$};
\node at (-6.5,1) {(2)};
\node at (-6,0) {$\wh\pi(q_{L_1})(\gamma)~~\times$};
\end{scope}
\begin{scope}[shift={(0,-3)}]
\draw [thick](-4,0) -- (-2,0);
\node at (-4.5,0) {$L_1$};
\node at (0.5,0) {$L_2$};
\node[red] at (-2,-0.5) {$O$};
\node at (-5,1) {(3)};
\draw [thick](0,0) -- (-2,0);
\draw [ultra thick,white](-1.4,0) -- (-1.2,0);
\draw [thick,blue] (-1,0) ellipse (0.3 and 0.7);
\draw [ultra thick,white](-0.7,-0.1) -- (-0.7,0.1);
\draw [red,fill=red] (-2,0) node (v1) {} ellipse (0.05 and 0.05);
\draw [dashed,thick](-0.7396,0) -- (-0.6396,0);
\node[blue] at (-1,-1) {$\gamma\in\Gamma^{(1)}$};
\end{scope}
\begin{scope}[shift={(8.8,-3)}]
\draw [thick](-4,0) -- (-2,0);
\node at (-4.5,0) {$L_1$};
\draw [thick](0,0) -- (-2,0);
\draw [red,fill=red] (-2,0) node (v1) {} ellipse (0.05 and 0.05);
\node at (0.5,0) {$L_2$};
\node[red] at (-2,-0.5) {$O$};
\node at (-6.5,1) {(4)};
\node[] at (-6,0) {$\wh\pi(q_{L_2})(\gamma)~~\times$};
\end{scope}
\end{tikzpicture}}
\caption{Suppose there exists a local operator $O$ between two line defects $L_1$ and $L_2$. Consider the correlation function (1), in which we have linked $L_1$ by a topological codimension-two defect $\gamma\in\Gamma^{(1)}$. Squeezing $\gamma$ on top of $L_1$, we obtain correlation function (2) with an extra phase factor $\wh\pi(q_{L_1})(\gamma)\in U(1)$. The correlation function (1) is equal to the correlation function (3), in which $\gamma$ now links $L_2$. Squeezing $\gamma$ on top of $L_2$, we obtain correlation function (4) with an extra phase factor $\wh\pi(q_{L_2})(\gamma)\in U(1)$. For consistency we must have $\wh\pi(q_{L_1})=\wh\pi(q_{L_2})$.}
\label{conscond}
\end{figure}
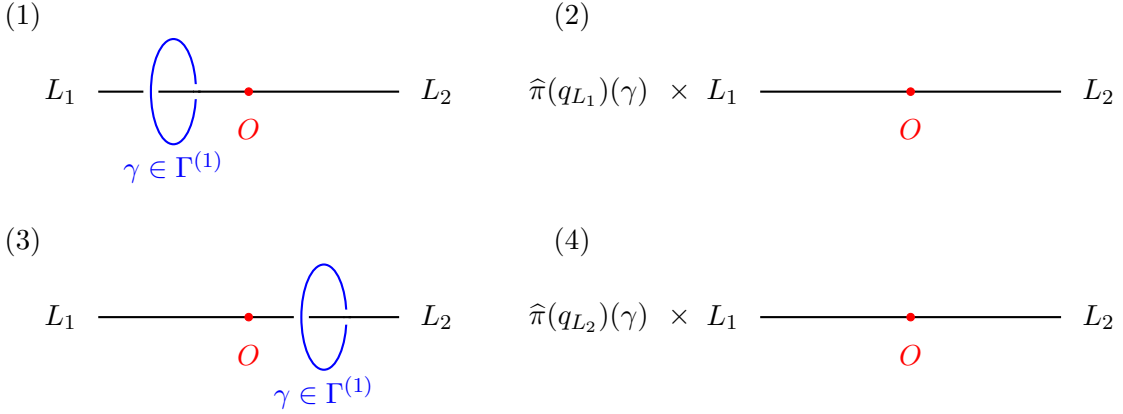

\paragraph{Charged objects and Bockstein 2-group symmetry.}
To each genuine line defect $L$, we can associate an element
\be
q_L\in\wh\cE\,,
\ee
where $\wh\cE$ is the Pontryagin dual of $\cE$. This element is such that
\be
\wh\pi(q_L)\in\wh\Gamma^{(1)}\,,
\ee
describes the charge of $L$ under the 1-form symmetry $\Gamma^{(1)}$. Here
\be
\wh\pi:~\wh\cE\to\wh\Gamma^{(1)}\,,
\ee
is the projection map dual to the injection map
\be\label{eq9}
i:~\Gamma^{(1)}\to\cE\,,
\ee
appearing in the short exact sequence (\ref{sesN}). 
Two line defects $L_1$ and $L_2$ such that
\be
\wh\pi(q_{L_1})\neq\wh\pi(q_{L_2})\,,
\ee
cannot be connected together by a local operator, because this would be in violation of charge conservation for 1-form symmetry $\Gamma^{(1)}$. See figure \ref{conscond}. Instead, let's assume that we have picked line defects $L_1$ and $L_2$ such that
\be\label{eq8}
\wh\pi(q_{L_1})=\wh\pi(q_{L_2})\,.
\ee
Moreover, suppose that there exists a local operator $O$ that can convert $L_1$ into $L_2$ as in figure \ref{conscond}. Then, $O$ must live in a representation $R$ of $F$ such that
\be
q_R=q_{L_2}-q_{L_1}\in\wh\cZ\,,
\ee
where $q_R\in\wh\cZ$ is the charge of $R$ under $\cZ$ as defined in (\ref{eq7}), and the fact that $q_{L_2}-q_{L_1}$ is in $\wh\cZ$ subgroup of $\wh\cE$ is seen by combining the identity (\ref{eq8}), with the fact that we have a short exact sequence
\be
0\to\wh\cZ\to\wh\cE\to\wh\Gamma^{(1)}\to0\,,
\ee
which is Pontryagin dual sequence of the short exact sequence (\ref{sesN}).

\paragraph{Backgrounds for Bockstein 2-group symmetry.}
A background for a Bockstein 2-group symmetry comprises of the following information:
\ben
\item A map
\be
A_\cF:~M_d\to B\cF\,,
\ee
using which we can pull-back the principal $\cF$-bundle $E\cF$ on $B\cF$ to a principal $\cF$-bundle $A_\cF^*(E\cF)$ on $M_d$.
\item A connection on $A_\cF^*(E\cF)$.
\item A background field
\be
B_w\in Z^2(M_d,\cE)\,,
\ee
which is an $\cE$-valued 2-cocycle on $M_d$ such that
\be
\pi_*B_w=A_\cF^*w_2\,,
\ee
where
\be
\pi_*:~Z^2(\cdot,\cE)\to Z^2(\cdot,\cZ)\,,
\ee
is the push-forward map associated to the projection map
\be\label{eq10}
\pi:~\cE\to\cZ\,,
\ee
appearing in the short exact sequence (\ref{sesN}), and $A_\cF^*$ is the pull-back map associated to $A_\cF$.
\een

\paragraph{1-form background from Bockstein 2-group background.}
Given a choice of
\be
\wt w_2\in C^2(B\cF,\cE)\,,
\ee
which is an $\cE$-valued 2-cochain on $B\cF$ lifting $w_2$ in the sense that
\be
\pi_*\wt w_2=w_2\,,
\ee
we can express $B_w$ as
\be
B_w=i_*B_2+A_\cF^*\wt w_2\,,
\ee
where
\be
i_*:~C^2\left(\cdot,\Gamma^{(1)}\right)\to C^2(\cdot,\cE)\,,
\ee
is the push-forward map associated to the injection map (\ref{eq9}), and
\be
B_2\in C^2\left(M_d,\Gamma^{(1)}\right)\,,
\ee
is a $\Gamma^{(1)}$-valued 2-cochain on $M_d$, which is identified with the background field for the 1-form symmetry group $\Gamma^{(1)}$. Using the closedness of $B_w$, one can now show that the defining relation (\ref{B2g}) for a Bockstein 2-group symmetry is satisfied:
\begin{align}
\delta B_w&=i_*\delta B_2+A_\cF^*i_*\text{Bock}(w_2)\,,\\
0&=i_*\big(\delta B_2+A_\cF^*\text{Bock}(w_2)\big)\,,\\
0&=\delta B_2+A_\cF^*\text{Bock}(w_2)\,.
\end{align}

\subsection{Disconnected 2-Group Symmetry}
\label{sec:discconn2groupsym}
Let's now include a finite 0-form symmetry group $\Gamma^{(0)}$ and study its interaction with the connected 2-group symmetry discussed in the previous subsection.

\paragraph{General disconnected 2-group symmetries.}
A general disconnected 2-group symmetry is a tuple
\be\label{d2gG}
\wt\fG=\left(\Gamma^{(1)},\wt\cF,\rho,[\wt\Theta]\right)\,,
\ee
with ingredients defined as follows. $\Gamma^{(1)}$ is the 1-form symmetry group and the semi-direct product
\be
\wt\cF=\cF\rtimes\Gamma^{(0)}\,,
\ee
is the full disconnected 0-form symmetry group. The map
\be
\rho:~\Gamma^{(0)}\to\text{Aut}\left(\Gamma^{(1)}\right)\,,
\ee
captures the action of finite part $\Gamma^{(0)}$ of the 0-form symmetry on the 1-form symmetry, and
\be
[\wt\Theta]\in H^3_{w_1}\left(B\wt\cF,\Gamma^{(1)}\right)\,,
\ee
which we call the \textit{twisted Postnikov class} associated to the disconnected 2-group symmetry, is a $\Gamma^{(1)}$-valued class in $w_1$-twisted cohomology of $B\wt\cF$ defined using the action $\rho$ of $\Gamma^{(0)}$ on $\Gamma^{(1)}$. Here
\be
w_1\in Z^1\left(B\wt\cF,\Gamma^{(0)}\right)\,,
\ee
is a fixed representative of the characteristic class $[w_1]\in H^1\left(B\wt\cF,\Gamma^{(0)}\right)$ capturing the obstruction for restricting $\wt\cF$ bundles to $\cF$ bundles.

Physically, the twisted Postnikov class encodes the non-closedness of the background field $B_2$ for 1-form symmetry via
\be
\delta_{B_1} B_2+A_{\wt\cF}^*\wt\Theta=0\,,
\ee
where $\delta_{B_1}$ is a twisted differential defined using the background field 
\be
B_1=A_{\wt\cF}^*w_1\in Z^1\left(M_d,\Gamma^{(0)}\right)\,,
\ee
for the finite 0-form symmetry $\Gamma^{(0)}$ and the action $\rho$, where
\be
\wt\Theta\in Z^3_{w_1}\left(B\wt\cF,\Gamma^{(1)}\right)\,,
\ee
is a fixed $\Gamma^{(1)}$-valued $w_1$-twisted 3-cocycle on $B\wt\cF$ lying in the class $[\wt\Theta]$ and the twist is again defined using $\rho$.

\paragraph{Connected part of $\wt\fG$.}
The connected part of a disconnected 2-group $\wt\fG$ is a connected 2-group $\fG$ of the form
\be
\fG=\left(\Gamma^{(1)},\cF,[\Theta]\right)\,,
\ee
where $\cF$ is the connected part of the disconnected 0-form symmetry group $\wt\cF$, and the Postnikov class $[\Theta]$ is obtained from the twisted Postnikov class $[\wt\Theta]$ via
\be
[\Theta]=i_\cF^*[\wt\Theta]\,,
\ee
where $i_\cF^*$ is the pullback map
\be
i_\cF^*:~H^3_{w_1}\left(B\wt\cF,\Gamma^{(1)}\right)\to H^3\left(B\cF,\Gamma^{(1)}\right)\,,
\ee
associated to the inclusion map
\be
i_\cF:~\cF\to\wt\cF\,.
\ee
The existence of such a pullback map relies crucially on the fact that twist is performed using the class $[w_1]$ relating $\cF$ and $\wt\cF$.

\paragraph{Disconnected Bockstein 2-group symmetries.}
In this paper, we focus on particular types of disconnected 2-group symmetries, which we call ``disconnected Bockstein 2-group symmetries'', for which we have the relationship
\be\label{Bd2g}
\delta_{B_1}B_2+A_{\wt\cF}^*\text{Bock}_{w_1}(w_2)=0\,,
\ee
where
\be
w_2\in Z^2_{w_1}(B\wt\cF,\cZ)\,,
\ee
is a $w_1$-twisted cocycle lying in the characteristic class
\be
[w_2]\in H^2_{w_1}(B\cF,\cZ)\,,
\ee
capturing the obstruction for lifting $\wt\cF$ bundles to $\wt F$ bundles. That is, for a disconnected Bockstein 2-group symmetry, the Postnikov class can be expressed as
\be
[\wt\Theta]=\text{Bock}_{w_1}([w_2])\,,
\ee
where
\be
\text{Bock}_{w_1}:~H^2_{w_1}\left(B\wt\cF,\cZ\right)\to H^3_{w_1}\left(B\wt\cF,\Gamma^{(1)}\right)\,,
\ee
is the $w_1$-twisted Bockstein homomorphism\footnote{See appendix \ref{sec:bockrev} for definition.} in the long exact sequence for $w_1$-twisted cohomologies associated to a short exact sequence
\be\label{sesN2}
0\to\Gamma^{(1)}\to\cE\to\cZ\to0\,,
\ee
with an action of $\Gamma^{(0)}$ on $\cE$, which when restricted to the subgroup $\Gamma^{(1)}\subseteq\cE$ becomes the action $\rho$. Such an action of $\Gamma^{(0)}$ on $\cE$ projects to a well-defined action of $\Gamma^{(0)}$ on $\cZ$.

It can be checked that the connected part of a disconnected Bockstein 2-group symmetry associated to the short exact sequence (\ref{sesN2}) along with a suitable action of $\Gamma^{(0)}$ on $\cE$, is a connected Bockstein 2-group symmetry associated to the same short exact sequence (\ref{sesN2}).

\paragraph{Charged objects and disconnected Bockstein 2-group symmetry.}
Consider a genuine line defect $L$, which is transformed to a genuine line defect $o\cdot L$ under an element $o\in\Gamma^{(0)}$. Then, we have
\be
q_{o\cdot L}=o\cdot q_L\in\wh\cE\,,
\ee
where $q_{o\cdot L}$ is the element of $\wh\cE$ associated to the line defect $o\cdot L$, and $o\cdot q_L$ is the element obtained by applying the action of $\Gamma^{(0)}$ on $\wh\cE$ to the element $q_L\in\wh\cE$. This action of $\Gamma^{(0)}$ on $\wh\cE$ is obtained as the dual of the action of $\Gamma^{(0)}$ on $\cE$.

Now, consider a local operator $O$ that can convert a line defect $L_1$ into another line defect $L_2$. Then, as we discussed in the previous subsection, this is possible only if
\be
q_{L_2}-q_{L_1}\in\wh\cZ\subseteq\wh\cE\,,
\ee
and moreover $O$ must transform in a representation $R$ of $F$ such that
\be
q_R=q_{L_2}-q_{L_1}\,,
\ee
is the charge of $R$ under $\cZ$. Now, under $o\in\Gamma^{(0)}$, $O$ is sent to a local operator $o\cdot O$ that can convert the line defect $o\cdot L_1$ into the line defect $o\cdot L_2$. Moreover, the representation of $F$ formed by $o\cdot O$ is simply the representation $o\cdot R$ obtained by applying the action of $o$ on $R$, which indeed has the property that
\be
q_{o\cdot R}=q_{o\cdot L_2}-q_{o\cdot L_1}\in\wh\cZ\subseteq\wh\cE\,.
\ee

\paragraph{Backgrounds for disconnected Bockstein 2-group symmetry.}
A background for a disconnected Bockstein 2-group symmetry comprises of the following information:
\ben
\item A map
\be
A_{\wt\cF}:~M_d\to B\wt\cF\,,
\ee
using which we can pull-back the principal $\wt\cF$-bundle $E\wt\cF$ on $B\wt\cF$ to a principal $\wt\cF$-bundle $A_{\wt\cF}^*(E\wt\cF)$ on $M_d$.\\
This fixes the background field $B_1$ for the finite 0-form symmetry via
\be
B_1=A_{\wt\cF}^*w_1\,.
\ee
\item A connection on $A_{\wt\cF}^*(E\wt\cF)$.
\item A background field
\be
B_w\in Z^2_{B_1}(M_d,\cE)\,,
\ee
which is an $\cE$-valued $B_1$-twisted 2-cocycle on $M_d$ such that
\be
\pi_*B_w=A_{\wt\cF}^*w_2\,.
\ee
\een

\paragraph{1-form background from disconnected Bockstein 2-group background.}
Given a choice of
\be
\wt w_2\in C^2(B\cF,\cE)\,,
\ee
which is an $\cE$-valued 2-cochain on $B\cF$ lifting $w_2$ in the sense that
\be
\pi_*\wt w_2=w_2\,,
\ee
we can express $B_w$ as
\be
B_w=i_*B_2+A_{\wt\cF}^*\wt w_2\,,
\ee
where
\be
B_2\in C^2\left(M_d,\Gamma^{(1)}\right)\,,
\ee
is a $\Gamma^{(1)}$-valued 2-cochain on $M_d$, which is identified with the background field for 1-form symmetry group $\Gamma^{(1)}$. Using the twisted-closedness of $B_w$, one can now show that the defining relation (\ref{Bd2g}) for a disconnected Bockstein 2-group symmetry is satisfied:
\begin{align}
\delta_{B_1} B_w&=i_*\delta_{B_1} B_2+A_{\wt\cF}^*i_*\text{Bock}_{w_1}(w_2)\,,\\
0&=i_*\big(\delta_{B_1} B_2+A_{\wt\cF}^*\text{Bock}_{w_1}(w_2)\big)\,,\\
0&=\delta_{B_1} B_2+A_{\wt\cF}^*\text{Bock}_{w_1}(w_2)\,.
\end{align}

\paragraph{A special case.} 
We now consider circumstances where the twisted Bockstein Postnikov class reduces to a simpler expression. Consider the special case where the action of $\Gamma^{(0)}$ on $\Gamma^{(1)}$ and $\cZ$ is trivial, but the action on $\cE$ is non-trivial. It should be noted that even though $\Gamma^{(0)}$ has trivial action on $\cZ$, it may still have a non-trivial action on the connected 0-form symmetry group $\cF$. In such a situation, we claim that the twisted Postnikov class can be described as
\be\label{PCSC}
[\wt\Theta]=\Bock([w_2])+[w_1]\cup [w_2] \,.
\ee
Here $[w_2]\in H^2(B\wt\cF,\cZ)$ is, as before, the characteristic class describing the obstruction of lifting $\wt\cF$ bundles to $\wt F$ bundles, but it is now a non-twisted class because the action of $\Gamma^{(0)}$ on $\cZ$ is trivial. Moreover, the term $\Bock([w_2])$ is a non-twisted Bockstein homomorphism associated to the short exact sequence (\ref{sesN2}). In other words, in such a situation, the twisted Bockstein becomes the sum of a non-twisted Bockstein and a cup product with $w_1$. 

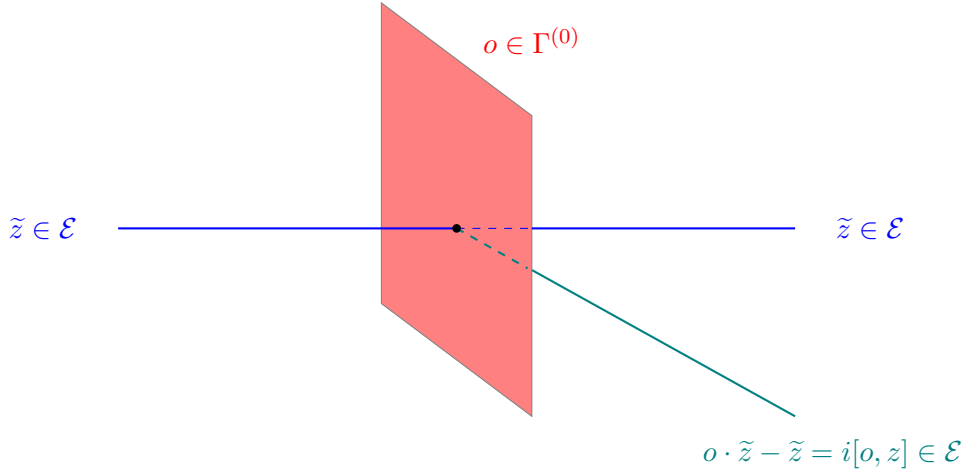
\begin{figure}
\centering
\begin{tikzpicture}
\draw[fill=red,opacity=0.5] (-1,1) -- (-1,2.5) -- (1,1) -- (1,-3) -- (-1,-1.5)--(-1,1);
\draw[thick,blue] (-4.5,-0.5) -- (0,-0.5);
\draw [thick,blue] (4.5,-0.5) -- (1,-0.5);
\draw [blue,dashed](0,-0.5) -- (1,-0.5);
\node[blue] at (-5.5,-0.5) {$\wt z\in\cE$};
\node[blue] at (5.5,-0.5) {$\wt z\in\cE$};
\node[red] at (1,2) {$o\in\Gamma^{(0)}$};
\draw [teal,thick](1.002,-1.0548) -- (4.5,-3);
\draw [teal,thick,dashed](0,-0.5) -- (0.9442,-1.0311);
\draw [black,fill=black] (0,-0.5) ellipse (0.05 and 0.05);
\node [teal]at (5,-3.5) {$o\cdot\wt z-\wt z=i[o,z]\in\cE$};
\end{tikzpicture}
\caption{Pictorial representation of the homomorphism (\ref{hom}). In blue and teal are shown pieces of $(d-2)$-cycle Poincare dual to the $\cE$ valued 2-cocycle $B_w$. The blue piece is labeled by an $\wt z\in\cE$ which is a lift of $z\in\cZ$. Upon passing the blue piece through codimension-one topological defect $o\in\Gamma^{(0)}$, we obtain an additional piece, shown in teal, labeled by the element $i[o,z]$ living in the $\Gamma^{(1)}$ subgroup of $\cE$.}
\label{pichom}
\end{figure}

The cup product appearing in (\ref{PCSC}) is defined via a bi-homomorphism
\be\label{hom}
[\cdot,\cdot]:\Gamma^{(0)}\times\cZ\to\Gamma^{(1)}\,,
\ee
which is constructed as follows. Pick elements $o\in\Gamma^{(0)}$ and $z\in\cZ$. Let $\wt z\in\cE$ be a lift of $z$. By the triviality of the action of $\Gamma^{(0)}$ on $\cZ$, we know that the element $o\cdot\wt z\in\cE$ obtained by acting $o$ on $\wt z$ is also a lift of $z$. Thus, $o\cdot\wt z-\wt z\in i\left(\Gamma^{(1)}\right)$ and we define
\be
[o,z]=i^{-1}\big(o\cdot\wt z-\wt z\big)\in\Gamma^{(1)}\,.
\ee
To show that the above map does not depend on the lift $\wt z$ of $z$, let us consider another lift $\wt z'\in\cE$ of $z$, which can be expressed as
\be
\wt z'=\wt z+i(\gamma) \,,
\ee
for some $\gamma$ in $\Gamma^{(1)}$. Now we have
\be
o\cdot\wt z'=o\cdot\wt z+i(o\cdot\gamma)=o\cdot\wt z+i(\gamma)\,,
\ee
where we have used the fact that the action of $\Gamma^{(0)}$ on $\Gamma^{(1)}$ is trivial. Thus, we have
\be
o\cdot\wt z'-\wt z'=o\cdot\wt z-\wt z\,.
\ee
The fact that (\ref{hom}) defined this way is a group homomorphism from the point of view of the $\cZ$ factor follows trivially by using a lift $\wt z_1+\wt z_2$ of $z_1+z_2$, where $\wt z_i$ is a lift of $z_i$. On the other hand, to show that (\ref{hom}) defined this way is a group homomorphism from the point of view of the $\Gamma^{(0)}$ factor, compute $[o_1,z]$ using a general lift $\wt z$, but compute $[o_2,z]$ using the lift $o_1\cdot\wt z+i\left([o_1,z]\right)$. Then, it follows from a direct computation that
\be
[o_1,z]+[o_2,z]=i^{-1}\big(o_1\cdot\wt z-\wt z+o_2\cdot\left(o_1\cdot\wt z+i([o_1,z])\right)-o_1\cdot\wt z-i([o_1,z])\big)=[o_2o_1,z]\,.
\ee
See figure \ref{pichom} for a pictorial interpretation of the homomorphism (\ref{hom}).

The derivation of the expression (\ref{PCSC}) for twisted Postnikov class can be found in appendix \ref{sec:bockrev}.

If, in addition, the short exact sequence (\ref{sesN2}) splits, the twisted Postnikov class further simplifies to
\be
[\wt\Theta]=[w_1]\cup[w_2] \,,
\ee
as untwisted Bockstein homomorphism is trivial in such a situation. The Postnikov class for the connected part of 2-group is obtained by simply setting $[w_1]=0$, implying that the connected part of 2-group is trivial. Thus, we see that a disconnected 2-group can be non-trivial, even if its connected component is trivial!

\subsection{2-Group Symmetries in Gauge Theories}
In this subsection, we describe how disconnected 2-group symmetries arise in gauge theories of the type being considered in this paper.

\paragraph{Connected part of 2-group symmetry.}
The electric 1-form symmetry group $\Gamma^{(1)}$ and the connected part $\cF$ of flavor 0-form symmetry group can combine to form a Bockstein 2-group symmetry. 

Let us review the construction of the group $\cE$ and the short exact sequence (\ref{sesN}). We begin by collecting the charges under $Z_\cG\times Z_F$ of all the matter fields in the gauge theory. These charges form a set $M$ of $\wh Z_\cG\times\wh Z_F$. Let
\be
\cM:=\text{Span}(M)\subseteq\wh Z_\cG\times\wh Z_F\,,
\ee
be the subgroup of $\wh Z_\cG\times\wh Z_F$ generated by $M$. Then, we can describe the Pontryagin dual $\wh\cE$ of $\cE$ as
\be
\wh\cE=\frac{\wh Z_\cG\times\wh Z_F}{\cM}\,.
\ee
Dually, $\cE$ is the subgroup of $Z_\cG\times Z_F$ that acts trivially on the set $M$ of charges.

Since the group $\Gamma^{(1)}$ of electric 1-form symmetries is defined as the subgroup of $Z_\cG$ leaving $M$ invariant, we can recognize $\Gamma^{(1)}$ as the following subgroup of $\cE$
\be
\Gamma^{(1)}=Z_\cG\cap\cE\subseteq\cE\,.
\ee
This provides us with the inclusion map $i:~\Gamma^{(1)}\to\cE$ in the short exact sequence (\ref{sesN}). On the other hand $\cZ$ is the subgroup of $Z_F$ that leaves
\be
M_F=M\cap\wh Z_F\,,
\ee
invariant. Thus we can recognize $\cZ$ as
\be
\cZ=\text{Im}_{\pi_F}(\cE)\subseteq Z_F\,,
\ee
which is the image of $\cE$ under the obvious projection map
\be
\pi_F:~Z_\cG\times Z_F\to Z_F\,.
\ee
This provides the projection map $\pi:~\cE\to\cZ$ in the short exact sequence (\ref{sesN}). The reader can easily check that the maps $i$ and $\pi$ defined this way indeed sit in a short exact sequence.

\paragraph{Disconnected 2-group symmetry.}
Now let us include outer-automorphism symmetry group $\Gamma^{(0)}\subseteq\cO_\fg\times\cO_\ff$. As discussed in previous sections, the elements of $\Gamma^{(0)}$ act on $\wh Z_\cG$ and $\wh Z_F$ via their projections $\pi_\fg\left(\Gamma^{(0)}\right)\subseteq\cO_\fg$ and $\pi_\ff\left(\Gamma^{(0)}\right)\subseteq\cO_\ff$ respectively. Combining the two actions, we obtain an action of $\Gamma^{(0)}$ on $\wh Z_\cG\times\wh Z_F$.

Since we must have a closed action of $\Gamma^{(0)}$ on the subset $M$ of $\wh Z_\cG\times\wh Z_F$, we learn that the action of $\Gamma^{(0)}$ on $\wh Z_\cG\times\wh Z_F$ projects to a well-defined action of $\Gamma^{(0)}$ on $\wh\cE$, from which we obtain the action of $\Gamma^{(0)}$ on $\cE$ required to fully determine the disconnected Bockstein 2-group symmetry of the gauge theory.

\paragraph{Connected part of structure group.}
As we have discussed in previous sections, the electric 1-form symmetry group $\Gamma^{(1)}$ modifies the allowed gauge bundles that a gauge theory sums over, and the connected part $\cF$ of the 0-form symmetry group describes the allowed bundles for continuous 0-form symmetry backgrounds. In general, gauge and 0-form bundles combine to form 
bundles for the Lie group
\be
\cS:=\frac{\cG\times F}{\cE}\,,
\ee
known as the \textit{structure group} associated to the gauge theory. As finite 0-form symmetries are not included, $\cS$ should more precisely be referred to as the connected part of the full structure group.

In the presence of a background for Bockstein 2-group symmetry, the gauge theory sums over $\cS$ bundles 
\be
A_\cS:~M_d\to\cS\,,
\ee
such that
\be
\pi_\cS\circ A_\cS=A_\cF\,,
\ee
and
\be
B_w=A_\cS^*w_2^\cE\,.
\ee
Here
\be
\pi_\cS:~\cS\to\cF\,,
\ee
is the natural projection map and
\be
w_2^\cE\in Z^2(B\cS,\cE)\,,
\ee
is a fixed representative of the characteristic class $\left[w_2^\cE\right]\in H^2(B\cS,\cE)$ capturing the obstruction for lifting $\cS$ bundles to $\cG\times F$ bundles.

\paragraph{Disconnected structure group.}
Including the finite 0-form symmetry group $\Gamma^{(0)}$, we can write the full disconnected structure group $\wt\cS$ as
\be
\wt\cS=\cS\rtimes\Gamma^{(0)}=\frac{\cG\times F}{\cE}\rtimes\Gamma^{(0)}=\frac{(\cG\times F)\rtimes\Gamma^{(0)}}{\cE}\,,
\ee
where the action of $\Gamma^{(0)}$ on $\cS$ is obtained by projecting the action of $\Gamma^{(0)}$ on $\cG\times F$, which is well-defined because we have a closed action of $\Gamma^{(0)}$ on $\cE$.

In the presence of a background for disconnected Bockstein 2-group symmetry, the gauge theory sums over $\wt\cS$ bundles 
\be
A_{\wt\cS}:~M_d\to B\wt\cS\,,
\ee
such that
\be
\pi_{\wt\cS}\circ A_{\wt\cS}=A_{\wt\cF}\,,
\ee
and
\be
B_w=A_{\wt\cS}^*w_2^\cE\,.
\ee
Here
\be
\pi_{\wt\cS}:~B\wt\cS\to B\wt\cF\,,
\ee
is the map induced by the natural projection map
\be
\wt\cS\to\wt\cF\,,
\ee
and
\be
w_2^\cE\in Z^2_{w_1^{\wt\cS}}(B\wt\cS,\cE)\,,
\ee
is a fixed representative of the characteristic class $\left[w_2^\cE\right]\in H^2_{w_1^{\wt\cS}}(B\wt\cS,\cE)$ capturing the obstruction for lifting $\wt\cS$ bundles to $(\cG\times F)\rtimes\Gamma^{(0)}$ bundles, where
\be
w_1^{\wt\cS}\in Z^1\left(B\wt\cS,\Gamma^{(0)}\right)\,,
\ee
is a fixed representative of the characteristic class $\left[w_1^{\wt\cS}\right]\in H^1\left(B\wt\cS,\Gamma^{(0)}\right)$ capturing the obstruction for restricting $\wt\cS$ bundles to $\cS$ bundles. Note that $w_1^{\wt\cS}$ can be expressed as
\be
w_1^{\wt\cS}=\pi_{\wt\cS}^*w_1\,,
\ee
where $w_1$ is a representative of the characteristic class $[w_1]\in H^1\left(B\wt\cF,\Gamma^{(0)}\right)$.
Consequently, the background field for $\Gamma^{(0)}$ can be expressed as
\be
B_1=A_{\wt\cS}^*w_1^{\wt\cS}\,.
\ee

\paragraph{Example of special case.}
Consider a $\cG=\Spin(2n)$ gauge theory with $m$ complex scalars in vector representation, such that they transform in fundamental representation of a 0-form symmetry algebra $\ff=\sp(m)$, with $F=\Sp(m)$ whose center is $Z_F=\Z_2$. The gauge invariant local operators are uncharged under $Z_F$, so the connected part of 0-form symmetry group is
\be
\cF=\text{PSp}(m)=\Sp(m)/\Z_2\,,
\ee
and so 
\be
\cZ=\Z_2\,.
\ee
The outer-automorphism of the gauge algebra $\fg=\so(2n)$ provides a 
\be
\Gamma^{(0)}=\Z_2\,,
\ee
0-form symmetry of the gauge theory. Since it does not involve outer-automorphisms of the flavor algebra $\ff$, the full disconnected 0-form symmetry group is
\be
\wt\cF=\text{PSp}(m)\times\Z_2\,.
\ee
To discuss 1-form and 2-group symmetries, we need to divide our analysis into two cases $n=2N+1$ and $n=2N$ as the center $Z_\cG$ of the gauge group is sensitive to the parity of $n$.

Let us first consider the case $n=2N+1$. In this case we have $Z_\cG=\Z_4$, such that the $\Z_2$ subgroup of $Z_\cG$ acts trivially on the vector representation. Consequently, the 1-form symmetry group is
\be
\Gamma^{(1)}=\Z_2\,.
\ee
Now, to compute $\cE$, note that the generator of $Z_\cG=\Z_4$ acts on the scalar fields as a multiplication by $-1$, which can be cancelled by the action of the generator of $Z_F=\Z_2$. Thus, we have
\be
\cE=\Z_4\subset Z_\cG\times Z_F\,,
\ee
and the short exact sequence (\ref{sesN}) is the non-split sequence
\be
0\to\Z_2\to\Z_4\to\Z_2\to0\,.
\ee
Consequently, we have a 2-group symmetry with Postnikov class
\be
[\Theta]=\Bock([w_2])\,,
\ee
where 
\be
[w_2]\in H^2(B\Sp(m),\Z_2)\,,
\ee
is the characteristic class capturing the obstruction of lifting background $\text{PSp}(m)$ bundles to $\Sp(m)$ bundles.

Now let us include the $\Gamma^{(0)}=\Z_2$ outer-automorphism symmetry. It acts on $Z_\cG=\Z_4$ by sending each element to its inverse, but has no action on $Z_F=\Z_2$. Thus it acts on $\cE=\Z_4\subset Z_\cG\times Z_F$ in a closed fashion by sending each element to its inverse. However, it has no action on $\Gamma^{(1)}=\Z_2$ and $\cZ=\Z_2$. Thus, we are in the special case discussed around (\ref{PCSC}). Thus, in order to determine the twisted Postnikov class of the full disconnected 2-group symmetry, we need to compute the homomorphism (\ref{hom}). Pick the element $z=1\in\cZ=\Z_2=\{0,1\}$ and its lift $\wt z=1\in\cE=\Z_4=\{0,1,2,3\}$. The lift $\wt z$ is sent to $o\cdot\wt z=3\in\cE=\Z_4=\{0,1,2,3\}$ by the action of $o=1\in\Gamma^{(0)}=\Z_2=\{0,1\}$. Thus we have $o\cdot\wt z-\wt z=2\in\cE=\Z_4=\{0,1,2,3\}$ which is the image of $1\in\Gamma^{(1)}=\Z_2=\{0,1\}$. This means that the homomorphism $[\cdot,\cdot]:~\Z_2\times\Z_2\to\Z_2$ is multiplication modulo 2, if each $\Z_2$ is represented as an additive group with elements $\{0,1\}$. Thus, the twisted Postnikov class is
\be\label{cDc}
[\wt\Theta]=\Bock([w_2])+[w_1]\cup[w_2]\,,
\ee
where the cup product is the standard cup product on $\Z_2$ valued cohomology, and
\be
[w_1]\in H^1\big(B[\text{PSp}(m)\times \Z_2],\Z_2\big)\,,
\ee
is the characteristic class capturing the obstruction of restricting background $\text{PSp}(m)\times\Z_2$ bundles to $\text{PSp}(m)$ bundles.

The full disconnected structure group can be written as
\be
\wt\cS=\frac{\text{Pin}^+(4N+2)\times \Sp(m)}{\Z_4}=\frac{O(4N+2)\times \Sp(m)}{\Z_2}\,,
\ee
with its connected part being
\be
\cS=\frac{\text{Spin}(4N+2)\times \Sp(m)}{\Z_4}=\frac{SO(4N+2)\times \Sp(m)}{\Z_2}\,.
\ee
Now let us consider the case $n=2N$. In this case we have $Z_\cG=\Z_2\times\Z_2$, such that the $\Z_2$ diagonal subgroup of $Z_\cG$ acts trivially on the vector representation. Consequently, the 1-form symmetry group is
\be
\Gamma^{(1)}=\Z_2\,.
\ee
Now, to compute $\cE$, note that the generator of either of the $\Z_2$ factors inside $Z_\cG=\Z_2\times\Z_2$ acts on the scalar fields as a multiplication by $-1$, which can be cancelled by the action of the generator of $Z_F=\Z_2$. Thus, we have
\be
\cE=\Z_2\times\Z_2\subset Z_\cG\times Z_F\,,
\ee
where the generator of either of the $\Z_2$ factors of $\cE$ involves the generator of $Z_F=\Z_2$, while the generator of the diagonal $\Z_2$ in $\cE$ does not involve $Z_F$. The short exact sequence (\ref{sesN}) is the split sequence
\be
0\to\Z_2\to\Z_2\times\Z_2\to\Z_2\to0\,.
\ee
Consequently, we do not have a non-trivial connected 2-group symmetry.

Now let us include the $\Gamma^{(0)}=\Z_2$ outer-automorphism symmetry. It acts on $Z_\cG=\Z_2\times\Z_2$ by sending exchanging the two $\Z_2$ factors, but has no action on $Z_F=\Z_2$. Thus it acts on $\cE=\Z_2\times\Z_2\subset Z_\cG\times Z_F$ in a closed fashion by exchanging the two $\Z_2$ factors. However, it has no action on $\Gamma^{(1)}=\Z_2$ and $\cZ=\Z_2$. Thus, we are again in the special case discussed around (\ref{PCSC}). Since the sequence splits, the twisted Postnikov class is simply
\be\label{CdC}
[\wt\Theta]=[w_1]\cup[w_2]\,.
\ee
Thus, we have a non-trivial disconnected 2-group symmetry whose connected part does not involve any non-trivial mixing between 1-form and 0-form symmetries.

The full disconnected structure group can be written as
\be
\wt\cS=\frac{\text{Pin}^+(4N)\times \Sp(m)}{\Z_2\times\Z_2}=\frac{O(4N)\times \Sp(m)}{\Z_2}\,,
\ee
with its connected part being
\be
\cS=\frac{\text{Spin}(4N)\times \Sp(m)}{\Z_2\times\Z_2}=\frac{SO(4N)\times \Sp(m)}{\Z_2}\,.
\ee

\paragraph{Example of general case.}
Consider gauge group $\cG=SU(16)$ with 4 complex scalars transforming in the irrep $\L^4$ whose highest weight has Dynkin coefficients $(0,0,0,1,0,\cdots,0)$ and whose dimension is 1820. Let the algebra $\ff=\u(4)=\su(4)\oplus\u(1)$ rotating the four scalars describe a continuous 0-form symmetry of the theory. The center of $\cG$ is $Z_\cG=\Z_{16}$, whose $\Z_4$ subgroup is the 1-form symmetry
\be
\Gamma^{(1)}=\Z_4\,.
\ee
We have $F=U(4)$, with the 0-form symmetry group being
\be
\cF=U(4)/\Z_4\,,
\ee
and hence
\be
\cZ=\Z_4\subset Z_F=U(1)\,.
\ee
$\Gamma^{(1)}=\Z_4$ and $\cZ=\Z_4$ combine to form $\cE=\Z_{16}$ with the short exact sequence (\ref{sesN}) being the non-split sequence
\be
0\to\Z_4\to\Z_{16}\to\Z_4\to0\,,
\ee
whose associated Bockstein homomorphism describes the Postnikov class for the connected 2-group symmetry as
\be
[\Theta]=\Bock([w_2])\,,
\ee
where
\be
[w_2]\in H^2\left(B\frac{U(4)}{\Z_4},\Z_4\right)\,,
\ee
is the obstruction for lifting $U(4)/\Z_4$ bundles to $U(4)$ bundles. 

Now, we also have a
\be
\Gamma^{(0)}=\Z_2\,,
\ee
outer-automorphism symmetry that acts as the diagonal of the outer-automorphism of $\cG=SU(16)$ and $F=U(4)$. Thus, the full disconnected 0-form symmetry group is
\be
\wt\cF=\wt U(4)/\Z_4\,.
\ee
The reader can easily check that $\Gamma^{(0)}=\Z_2$ acts on $\Gamma^{(1)}$, $\cE$ and $\cZ$ by sending each element in either of these groups to its inverse in that group. Thus, we have a disconnected 2-group symmetry described by the twisted Postnikov class
\be
[\Theta]=\Bock_{w_1}([w_2])\,,
\ee
where
\be
[w_1]\in H^1\left(B\frac{\wt U(4)}{\Z_4},\Z_2\right)\,,
\ee
is the obstruction for restricting $\wt U(4)/\Z_4$ bundles to $U(4)/\Z_4$ bundles.

\subsection{Mixed 't Hooft Anomalies Dual to 2-Groups}
It is well-known that gauging a finite 1-form symmetry $\Gamma^{(1)}$ of a theory $\fT$ leads to a $(d-3)$-form symmetry
\be
\Gamma^{(d-3)}=\wh\Gamma^{(1)}\,,
\ee
in the gauged theory $\fT/\Gamma^{(1)}$.

\paragraph{Mixed anomaly dual to connected 2-group.}
Now consider a situation in which the 1-form symmetry participates in a connected 2-group of the form (\ref{2gG}) and moreover the 2-group does not suffer from any 't Hooft anomaly. This in particular means that the 1-form and 0-form components of the 2-group do not participate in any 't Hooft anomaly. Then, the dual $(d-3)$-form symmetry of the gauged theory $\fT/\Gamma^{(1)}$ has a mixed 't Hooft anomaly with the 0-form symmetry $\cF$ given by the anomaly theory
\be
\cA_{d+1}=\text{exp}\left(2\pi i\int_{M_{d+1}} A_\cF^*[\Theta]\cup[B_{d-2}]\right)\,,
\ee
where $[B_{d-2}]\in H^{d-2}\left(M_{d+1},\Gamma^{(d-3)}\right)$ is the background field for the $(d-3)$-form symmetry $\Gamma^{(d-3)}$, and the cup product is defined using the natural pairing $\wh\Gamma^{(1)}\times\Gamma^{(1)}\to\R/\Z$.

This can be shown beginning from the fact that the gauging is performed by adding to the Lagrangian a term of the form
\be
B_2\cup B_{d-2}\,,
\ee
and summing over $B_2$. 
Now performing a background gauge transformation
\be
B_{d-2}\to B_{d-2}+\delta\lambda_{d-3}\,,
\ee
the above term changes by
\be
\delta B_2\cup\lambda_{d-3}=A_\cF^*\Theta\cup\lambda_{d-3}\,,
\ee
which is the anomaly captured by the above anomaly theory $\cA_{d+1}$.

\paragraph{Mixed anomaly dual to disconnected 2-group.}
In a similar fashion, we can consider a situation in which the 1-form symmetry participates in a disconnected 2-group of the form (\ref{d2gG}) and moreover the 2-group does not suffer from any 't Hooft anomaly. 
Then, the dual $(d-3)$-form symmetry of the gauged theory $\fT/\Gamma^{(1)}$ has a mixed 't Hooft anomaly with the disconnected 0-form symmetry $\wt\cF$ given by the anomaly theory
\be
\wt\cA_{d+1}=\text{exp}\left(2\pi i\int_{M_{d+1}} A_{\wt\cF}^*[\wt\Theta]\cup[B_{d-2}]\right)\,,
\ee
where $[B_{d-2}]\in H_{B_1}^{d-2}\left(M_{d+1},\Gamma^{(d-3)}\right)$ is the background field for the $(d-3)$-form symmetry $\Gamma^{(d-3)}$. The action of $\Gamma^{(0)}$ on $\Gamma^{(d-3)}$ needed to define the above twisted cohomology is simply provided by the dual of the action of $\Gamma^{(0)}$ on $\Gamma^{(1)}$, as $\Gamma^{(d-3)}=\wh\Gamma^{(1)}$. Now, the cup product $A_{\wt\cF}^*[\wt\Theta]\cup[B_{d-2}]$ leads to an element of the untwisted cohomology $H^{d+1}(M_{d+1},\R/\Z)$.

This can be again be shown in a similar fashion as for the case of connected 2-groups discussed above. The gauging is performed by adding to the Lagrangian a term of the form
\be
B_2\cup B_{d-2}\,,
\ee
and summing over $B_2$. 
Now performing a background gauge transformation
\be
B_{d-2}\to B_{d-2}+\delta_{B_1}\lambda_{d-3}\,,
\ee
the above term changes by
\be
\delta_{B_1} B_2\cup\lambda_{d-3}=A_{\wt\cF}^*\wt\Theta\cup\lambda_{d-3}\,,
\ee
which is the anomaly captured by the above anomaly theory $\wt\cA_{d+1}$.

\paragraph{Examples.}
Consider the example of $\cG=\Spin(2n)$ gauge theory with $m$ complex scalars in vector representation that we discussed in the previous subsection. 

We saw that for $n=2N+1$, we have a disconnected 2-group symmetry with twisted Postnikov class (\ref{cDc}). After gauging 1-form symmetry $\Gamma^{(1)}=\Z_2$, we obtain a gauge theory with gauge group $G=SO(4N+2)$ with $m$ complex scalars in vector representation. According to the above analysis, there is a mixed anomaly in this theory given by
\be
\wt\cA_{d+1}=\text{exp}\left(2\pi i\int_{M_{d+1}} \Big(\Bock([w_2]+[w_1]\cup[w_2]\Big)\cup[B_{d-2}]\right)
\ee
where we have dropped the pull-back symbol $A_{\wt\cF}^*=A^*_{PSp(m)\times\Z_2}$ for improving clarity.

For $n=2N$, we have a disconnected 2-group symmetry with twisted Postnikov class (\ref{CdC}). After gauging 1-form symmetry $\Gamma^{(1)}=\Z_2$, we obtain a gauge theory with gauge group $G=SO(4N)$ with $m$ complex scalars in vector representation. According to the above analysis, there is a mixed anomaly in this theory given by
\be
\wt\cA_{d+1}=\text{exp}\left(2\pi i\int_{M_{d+1}} [w_1]\cup[w_2]\cup[B_{d-2}]\right)
\ee

\section*{Acknowledgements}
We thank Lea Bottini, Antoine Bourget, Sakura Schafer-Nameki and Jingxiang Wu for discussions. The work of LB is supported by ERC grants 682608 and 787185 under the European Union’s Horizon 2020 programme.

\appendix

\section{Lie Algebra Outer Automorphisms}
\label{sec:liealgeouterreview}
An automorphism $\Gamma$ of a Lie algebra $\mathfrak{g}$ is a map
\be
\Gamma: \mathfrak{g} \to \mathfrak{g}\,,
\ee
which preserves the commutation relations. The set of all automorphisms of a Lie algebra form a group Aut$(\mathfrak{g})$, where group multiplication is simply the composition of two maps. An automorphism of the form 
\be
\Gamma_y: x \in \mathfrak{g} \to [x,y] \in \mathfrak{g} \,,\quad \forall x \in \mathfrak{g}\,,
\ee
is called an \textit{inner} automorphism, the group of which is denoted Inn$(\mathfrak{g}) \subseteq \text{Aut}(\mathfrak{g})$. The quotient Aut$(\mathfrak{g})/$Inn$(\mathfrak{g})$ is called the \textit{outer}-automorphism group Out$(\mathfrak{g})$, which we also denote as $\cO_\fg$.

Suppose we begin with a simple, compact, non-abelian Lie algebra $\mathfrak{g}$ and its associated Dynkin diagram $\cD$. Invariance of the diagram $\cD$ under flips and rotations in the plane exactly capture the group of outer-automorphisms of the associated algebra. For example, Dynkin diagrams of type $\cD=A_{n}$ have a single flip symmetry. Their outer-automorphism group is $\Z_2$. $\cD=D_n$ diagrams have a $\Z_2$ exchange symmetry which swaps the two branches of the diagram. Furthermore, the algebra $\mathfrak{g}=\so(8)$ has a $\Z_3$ rotational symmetry as well as the $\Z_2$ symmetry it inherits from the branch exchange. The $\Z_3$ and $\Z_2$ symmetries do not commute, and combine to form a non-abelian symmetry group $S_3$, namely the group of permutations of three objects.

For a semi-simple, compact, non-abelian Lie algebra $\fg=\bigoplus_i\mathfrak{g}_i$, with Dynkin diagram $\cD=\bigsqcup_i\cD_i$ where $\cD_i$ is Dynkin diagram for $\fg_i$, the group $\cO_\fg$ of outer-automorphisms of $\fg$ can again be understood as symmetries of $\cD$, where now one can also exchange different Dynkin diagrams $\cD_i$ and $\cD_j$ if they are identical.

By its action on Dynkin diagram $\cD$, an outer-automorphism naturally acts on weights of $\fg$ by permuting the Dynkin coefficients of weights, and hence it acts naturally on representations of $\fg$.

\section{Semi-Direct Products and Disconnected Lie Groups}
\label{sec:semidirprod}
\paragraph{Semi-Direct Product Groups. } Semi-direct product groups are defined as follows. Consider a group $G$ and a group $\Gamma$. Given a homomorphism $\phi: \Gamma \to \text{Aut}(G)$, we construct the semi-direct product of $G$ and $\Gamma$, denoted $H_\phi = G \rtimes_{\phi} \Gamma$ out of elements $(g \in G, \gamma \in \Gamma) \in H_\phi$. The non-trivial structure in $H_\phi$ is in the group multiplication
\be
(g_1, \gamma_1) \cdot (g_2, \gamma_2) = (g_1 \phi(\gamma_1) \cdot g_2, \gamma_1 \gamma_2) \,,\quad \forall g_1,g_2 \in G\,, \gamma_1 \gamma_2 \in \Gamma\,.
\ee
\paragraph{Disconnected Lie Groups. }  In this work, we encounter semi-direct product groups in the context of disconnected Lie groups, which are Lie groups in which not every element can be continuously path connected to the identity element. We encounter disconnected Lie groups $\wt\cF$ of the form
\be
\wt\cF = \cF \rtimes \Gamma\,,
\ee
where $\cF$ is a connected Lie group and $\Gamma$ is a finite group acting on $\cF$.

\section{Twisted Cohomology}
\label{sec:twistedcoho}
In this section we review the modification of ordinary cohomology in the presence of an action by a finite $\Gamma^{(0)}$ on a coefficient group $G$. In particular, we review the twisted differential operation and the subsequent definition of twisted cohomology. We leave the manifold $X$ under consideration general. In our applications we will consider twisted cohomology on both spcaetime $X=M_d$ and the classifying space of $\wt\cF$ bundles $X=B\wt\cF$.

\paragraph{Orindary Differential.} The ordinary simplicial differential
\be
\delta: C^n(X,G) \to C^{n+1}(X,G) \,,
\ee
is defined through its action on simplicial n-cochains $f$
\be
(\delta f)_{i_0\dots i_{n+1}} = \sum_{j=0}^{j=n+1} (-1)^j f_{i_0 \dots \hat{i}_j \dots i_{n+1}} \,.
\ee
for some group $G$. We employ the notation that hatted indices are removed. 

\paragraph{Twisting. } Central to our work is the action of finite group $\Gamma^{(0)}$ on the group $G$. This action is described by a map
\be
\rho: \Gamma^{(0)} \to \Aut(G) \,.
\ee

\paragraph{Twisted Differential. } To account for such an action, one defines a twisted differential. We denote this as $\delta_{B_1}$, in terms of the background field $B_1$ which is a 1-cocycle taking values in $\Gamma^{(0)}$:
\be
(\delta_{B_1} f)_{i_0 \dots i_{n+1}} = \rho(B_{1,i_0 i_1}) f_{i_1 \dots i_n} + \sum_{j=1}^{j=n+1} (-1)^j f_{i_0 \dots \hat{i}_j \dots i_{n+1}} \,.
\ee
We can re-write $\delta_{B_1}$ in terms of the un-twisted version $\delta$ as 
\be
(\delta_{B_1} f)_{i_0 \dots i_{n+1}} = \rho(B_{1,i_0 i_1}) f_{i_1 \dots i_n} - f_{i_1 \dots i_n} + (\delta f)_{i_0 \dots i_{n+1}} \,.
\ee
Schematically we will write this as
\be
\delta_{B_1} f = \delta f + (\rho(B_1))-1)f \,.
\ee
For notational convenience, when we pick $X=B\wt\cF$, we denote the twisted differential operator $\delta_{w_1}$ 
\be
\delta_{w_1}: C^2(B\wt\cF, G) \to C^3(B\wt\cF,G) \,, 
\ee
which acts on cochains defined on the classifiying space of $\wt \cF$ bundles. As introduced earlier, 
\be
w_1 \in Z^1(B\wt\cF,\Gamma^{(0)})\,,
\ee
is a fixed representative of the characteristic class $[w_1] \in H^1(B\wt\cF,\Gamma^{(0)})$ capturing the obstruction for restricting $\wt\cF$ bundles to $\cF$ bundles. Given a map
\be
A_{\wt\cF}: M_d \to B\wt\cF \,,
\ee
the background field $B_1$ for $\Gamma^{(0)}$ is fixed in terms of this as 
\be
B_1 = A_{\wt\cF}^* w_1 \,.
\ee

\paragraph{Twisted Cohomology. } It is straightforward to show that the twisted differential is nilpotent. We can therefore define co-cycles and co-boundaries with respect to this new differential operator\footnote{These are called twisted co-cycles and twisted co-boundaries}. We will denote the group of twisted $p$ co-cycles as
\be
Z_{B_1}^p(X,\cdot) \,.
\ee
Therefore, using the twisted differential $\delta_{B_1}$, one can simply construct the \text{twisted} cohomology groups
\be
H_{B_1}^p(X,\cdot) \,,
\ee
analogously to ordinary cohomology (with the replacement $\delta \to \delta_{B_1}$). We also introduce the notation
\be
H_{w_1}^p(B\wt\cF,\cdot) \,,
\ee
for cohomology over the classifiying space of $\wt\cF$ bundles.

\section{(Twisted) Bockstein Homomorphism}
\label{sec:bockrev}
In this section we introduce the Bockstein homomorphism and its twisted cousin as well as providing a proof of its simplification in the special case introduced in section \ref{sec:discconn2groupsym}.

\paragraph{Bockstein construction.} We begin with a short exact sequence of the form
\be
1 \xrightarrow[]{i} \Gamma^{(1)} \to \cE \xrightarrow[]{\pi} \cZ \to 1 \,,
\ee
which appears throughout our 2-group constructions. There is an associated long exact sequence of the form
\be
\begin{tikzcd}
\cdots \rightarrow  H^2 (M_d, \Gamma^{(1)}) \arrow[r] & H^2 (M_d, \mathcal{E}) \arrow[r] 
\arrow[d, phantom, ""{coordinate, name=Z}] &  H^2 (M_d, \mathcal{Z}) \arrow[dll,
"\Bock",
rounded corners,
to path={ -- ([xshift=2ex]\tikztostart.east)
|- (Z) [near end]\tikztonodes
-| ([xshift=-3ex]\tikztotarget.west)
-- (\tikztotarget)}] \\
 H^3 (M_d, \Gamma^{(1)}) \arrow[r] & H^3 (M_d, \mathcal{E}) \arrow[r] &
 H^3 (M_d, \mathcal{Z}) \rightarrow \cdots \\
\end{tikzcd}
\ee
within which we have labelled the map
\be
\Bock: H^2 (M_d, \cZ) \to H^3 (M_d, \Gamma^{(1)}) \,.
\ee
This map is constructed as follows. Consider an equivalence class $[z] \in H^2(M,\cZ)$ and representative $z \in Z^2(M,\cZ)$. We lift to an element $\wt z \in C^n(M,\cE)$ such that $\pi(\wt z) = z$ and $\pi(\delta \wt z)=0$. Since $\delta \wt z$ is in the kernel of $\pi$, by exactness of the sequence we can write
\be
\delta \wt z = i(y) \,, 
\ee
for some $y \in C^3(M,\Gamma^{(1)})$. Since $i$ is injective, $\delta y = 0$ and one can show that $y \in H^3(M,\Gamma^{(1)})$. We therefore have constructed the Bockstein homomorphism via
\be
\Bock: [z] \in H^2(M_d,\cZ) \to [y] \in H^3(M_d, \Gamma^{(1)}) \,. 
\ee

\paragraph{Twisted Bockstein Homomorphism. } With the twisted cohomology definition above, we can define the \textit{twisted} Bockstein homomorphism 
\be
\text{Bock}_{w_1}:~H^2_{w_1}\left(B\wt\cF,\cZ\right)\to H^3_{w_1}\left(B\wt\cF,\Gamma^{(1)}\right)\,,
\ee
in the same way as the ordinary Bockstein homomorphism, now using a long exact sequence of twisted cohomologies
\be
\begin{tikzcd}
\cdots \rightarrow  H^2_{w_1} (M_d, \Gamma^{(1)}) \arrow[r] & H^2_{w_1} (M_d, \mathcal{E}) \arrow[r] 
\arrow[d, phantom, ""{coordinate, name=Z}] &  H^2_{w_1} (M_d, \mathcal{Z}) \arrow[dll,
"\Bock",
rounded corners,
to path={ -- ([xshift=2ex]\tikztostart.east)
|- (Z) [near end]\tikztonodes
-| ([xshift=-3ex]\tikztotarget.west)
-- (\tikztotarget)}] \\
 H^3_{w_1} (M_d, \Gamma^{(1)}) \arrow[r] & H^3_{w_1} (M_d, \mathcal{E}) \arrow[r] &
 H^3_{w_1} (M_d, \mathcal{Z}) \rightarrow \cdots \\
\end{tikzcd}
\ee
Crucially, this formalism captures the general case where $\Gamma^{(0)}$ can act on all three groups $\Gamma^{(1)}, \cE$ and $\cZ$ non-trivially.

\paragraph{Twisted Bockstein Homomorphism: A Special Case. }
We consider the special case where $\Gamma^{(0)}$ acts trivially on $\cZ, \Gamma^{(1)}$ but not $\cE$. Using the twisted differential $\delta_{w_1}$ we get a long exact sequence of the form
\be\label{eq:twistedLES}
\begin{tikzcd}
\cdots \rightarrow  H^2_{w_1} (B\cF, \Gamma^{(1)}) \arrow[r] & H^2_{w_1} (B\cF, \mathcal{E}) \arrow[r] 
\arrow[d, phantom, ""{coordinate, name=Z}] &  H^2_{w_1} (B\cF, \mathcal{Z}) \arrow[dll,
"\Bock_{w_1}",
rounded corners,
to path={ -- ([xshift=2ex]\tikztostart.east)
|- (Z) [near end]\tikztonodes
-| ([xshift=-3ex]\tikztotarget.west)
-- (\tikztotarget)}] \\
 H^3_{w_1} (B\cF, \Gamma^{(1)}) \arrow[r] & H^3_{w_1} (B\cF, \mathcal{E}) \arrow[r] &
 H^3_{w_1} (B\cF, \mathcal{Z}) \rightarrow \cdots \\
\end{tikzcd}
\ee
The twisting is trivial in the outside columns by construction:
\be
\begin{tikzcd}
\cdots \rightarrow  H^2 (B\cF, \Gamma^{(1)}) \arrow[r] & H^2_{w_1} (B\cF, \mathcal{E}) \arrow[r] 
\arrow[d, phantom, ""{coordinate, name=Z}] &  H^2 (B\cF, \mathcal{Z}) \arrow[dll,
"\Bock_{w_1}",
rounded corners,
to path={ -- ([xshift=2ex]\tikztostart.east)
|- (Z) [near end]\tikztonodes
-| ([xshift=-3ex]\tikztotarget.west)
-- (\tikztotarget)}] \\
 H^3 (B\cF, \Gamma^{(1)}) \arrow[r] & H^3_{w_1} (B\cF, \mathcal{E}) \arrow[r] &
 H^3 (B\cF, \mathcal{Z}) \rightarrow \cdots \\
\end{tikzcd}
\ee
We will now show that in this simplified situation, the twisted Bockstein homomorphism can be written as a sum of the \textit{un-twisted} Bockstein plus a cup product which we will define below:
\be
\Bock_{w_1} (z) = \Bock(z) + w_1 \cup z \,,
\ee
for some $z \in Z^2(B\wt\cF,\cZ)$.

First we consider the lift $\wt z \in C^2(B\cF,\cE)$ of $z \in Z^2(B\cF, \cZ)$. Using the above construction of the twisted Bockstein homomorphism, we know that
\be
\delta_{w_1} \wt z = i(y) \,,
\ee
for some $y\in C^3(B\cF,\Gamma^{(1)})$. Employing our decomposition of the twisted differential
\be\label{eq:bockdecomposition}
\delta_{w_1} \wt z = \delta \wt z + (\rho(w_1)-1) \wt z =  i(y) \,,
\ee
We will treat these two pieces in turn. We know from the above construction that we can write the first term as the un-twisted Bockstein homomorphism acting on $z$. Since $\rho$ acts trivially on $\cZ$ and $\Gamma^{(1)}$, both $\rho(w_1) \wt z$ and $\wt z$ are lifts of $z$ Therefore, $(\rho(w_1)-1) \wt z \in i(\Gamma^{(1)})$. This identity is use to define the bi-homomorphism
\be
[\cdot,\cdot]:\Gamma^{(0)}\times\cZ\to\Gamma^{(1)}\,,
\ee
which underpins the cup product $w_1 \cup z$. We therefore have shown
\be
\Bock_{w_1} (z) = \Bock(z) + w_1 \cup z \,.
\ee

\bibliographystyle{ytphys}
\small 
\baselineskip=.94\baselineskip
\let\bbb\bibitem\def\bibitem{\itemsep4pt\bbb}
\bibliography{ref}

\providecommand{\href}[2]{#2}\begingroup\raggedright\begin{thebibliography}{10}

\bibitem{Gaiotto:2014kfa}
D.~Gaiotto, A.~Kapustin, N.~Seiberg, and B.~Willett, ``{Generalized Global
  Symmetries},'' \href{http://dx.doi.org/10.1007/JHEP02(2015)172}{{\em JHEP}
  {\bfseries 02} (2015) 172}, \href{http://arxiv.org/abs/1412.5148}{{\ttfamily
  arXiv:1412.5148 [hep-th]}}.

\bibitem{Sharpe:2015mja}
E.~Sharpe, ``{Notes on generalized global symmetries in QFT},''
  \href{http://dx.doi.org/10.1002/prop.201500048}{{\em Fortsch. Phys.}
  {\bfseries 63} (2015) 659--682},
  \href{http://arxiv.org/abs/1508.04770}{{\ttfamily arXiv:1508.04770
  [hep-th]}}.

\bibitem{Tachikawa:2017gyf}
Y.~Tachikawa, ``{On gauging finite subgroups},''
  \href{http://dx.doi.org/10.21468/SciPostPhys.8.1.015}{{\em SciPost Phys.}
  {\bfseries 8} no.~1, (2020) 015},
  \href{http://arxiv.org/abs/1712.09542}{{\ttfamily arXiv:1712.09542
  [hep-th]}}.

\bibitem{Cordova:2018cvg}
C.~C\'ordova, T.~T. Dumitrescu, and K.~Intriligator, ``{Exploring 2-Group
  Global Symmetries},'' \href{http://dx.doi.org/10.1007/JHEP02(2019)184}{{\em
  JHEP} {\bfseries 02} (2019) 184},
  \href{http://arxiv.org/abs/1802.04790}{{\ttfamily arXiv:1802.04790
  [hep-th]}}.

\bibitem{Benini:2018reh}
F.~Benini, C.~C\'ordova, and P.-S. Hsin, ``{On 2-Group Global Symmetries and
  their Anomalies},'' \href{http://dx.doi.org/10.1007/JHEP03(2019)118}{{\em
  JHEP} {\bfseries 03} (2019) 118},
  \href{http://arxiv.org/abs/1803.09336}{{\ttfamily arXiv:1803.09336
  [hep-th]}}.

\bibitem{Hsin:2020nts}
P.-S. Hsin and H.~T. Lam, ``{Discrete Theta Angles, Symmetries and
  Anomalies},'' \href{http://dx.doi.org/10.21468/SciPostPhys.10.2.032}{{\em
  SciPost Phys.} {\bfseries 10} (2021) 032},
  \href{http://arxiv.org/abs/2007.05915}{{\ttfamily arXiv:2007.05915
  [hep-th]}}.

\bibitem{Cordova:2020tij}
C.~Cordova, T.~T. Dumitrescu, and K.~Intriligator, ``{2-Group Global Symmetries
  and Anomalies in Six-Dimensional Quantum Field Theories},''
  \href{http://arxiv.org/abs/2009.00138}{{\ttfamily arXiv:2009.00138
  [hep-th]}}.

\bibitem{Apruzzi:2021vcu}
F.~Apruzzi, L.~Bhardwaj, J.~Oh, and S.~Schafer-Nameki, ``{The Global Form of
  Flavor Symmetries and 2-Group Symmetries in 5d SCFTs},''
  \href{http://arxiv.org/abs/2105.08724}{{\ttfamily arXiv:2105.08724
  [hep-th]}}.

\bibitem{Bhardwaj:2021ojs}
L.~Bhardwaj, ``{Global Form of Flavor Symmetry Groups in 4d N=2 Theories of
  Class S},'' \href{http://arxiv.org/abs/2105.08730}{{\ttfamily
  arXiv:2105.08730 [hep-th]}}.

\bibitem{Bhardwaj:2021wif}
L.~Bhardwaj, ``{2-Group Symmetries in Class S},''
  \href{http://arxiv.org/abs/2107.06816}{{\ttfamily arXiv:2107.06816
  [hep-th]}}.

\bibitem{Lee:2021crt}
Y.~Lee, K.~Ohmori, and Y.~Tachikawa, ``{Matching higher symmetries across
  Intriligator-Seiberg duality},''
  \href{http://dx.doi.org/10.1007/JHEP10(2021)114}{{\em JHEP} {\bfseries 10}
  (2021) 114}, \href{http://arxiv.org/abs/2108.05369}{{\ttfamily
  arXiv:2108.05369 [hep-th]}}.

\bibitem{Apruzzi:2021mlh}
F.~Apruzzi, L.~Bhardwaj, D.~S.~W. Gould, and S.~Schafer-Nameki, ``{2-Group
  Symmetries and their Classification in 6d},''
  \href{http://arxiv.org/abs/2110.14647}{{\ttfamily arXiv:2110.14647
  [hep-th]}}.

\bibitem{DelZotto:2022joo}
M.~Del~Zotto, I.~n.~G. Etxebarria, and S.~Schafer-Nameki, ``{2-Group Symmetries
  and M-Theory},'' \href{http://arxiv.org/abs/2203.10097}{{\ttfamily
  arXiv:2203.10097 [hep-th]}}.

\bibitem{Cvetic:2022imb}
M.~Cveti\v{c}, J.~J. Heckman, M.~H\"ubner, and E.~Torres, ``{0-Form, 1-Form and
  2-Group Symmetries via Cutting and Gluing of Orbifolds},''
  \href{http://arxiv.org/abs/2203.10102}{{\ttfamily arXiv:2203.10102
  [hep-th]}}.

\bibitem{Bhardwaj:2021pfz}
L.~Bhardwaj, M.~Hubner, and S.~Schafer-Nameki, ``{1-form Symmetries of 4d N=2
  Class S Theories},'' \href{http://arxiv.org/abs/2102.01693}{{\ttfamily
  arXiv:2102.01693 [hep-th]}}.

\bibitem{Nguyen:2021naa}
M.~Nguyen, Y.~Tanizaki, and M.~\"Unsal, ``{Noninvertible 1-form symmetry and
  Casimir scaling in 2D Yang-Mills theory},''
  \href{http://dx.doi.org/10.1103/PhysRevD.104.065003}{{\em Phys. Rev. D}
  {\bfseries 104} no.~6, (2021) 065003},
  \href{http://arxiv.org/abs/2104.01824}{{\ttfamily arXiv:2104.01824
  [hep-th]}}.

\bibitem{Heidenreich:2021xpr}
B.~Heidenreich, J.~McNamara, M.~Montero, M.~Reece, T.~Rudelius, and
  I.~Valenzuela, ``{Non-Invertible Global Symmetries and Completeness of the
  Spectrum},'' \href{http://dx.doi.org/10.1007/JHEP09(2021)203}{{\em JHEP}
  {\bfseries 09} (2021) 203}, \href{http://arxiv.org/abs/2104.07036}{{\ttfamily
  arXiv:2104.07036 [hep-th]}}.

\bibitem{Apruzzi:2021phx}
F.~Apruzzi, M.~van Beest, D.~S.~W. Gould, and S.~Sch\"afer-Nameki,
  ``{Holography, 1-form symmetries, and confinement},''
  \href{http://dx.doi.org/10.1103/PhysRevD.104.066005}{{\em Phys. Rev. D}
  {\bfseries 104} no.~6, (2021) 066005},
  \href{http://arxiv.org/abs/2104.12764}{{\ttfamily arXiv:2104.12764
  [hep-th]}}.

\bibitem{Hosseini:2021ged}
S.~S. Hosseini and R.~Moscrop, ``{Maruyoshi-Song Flows and Defect Groups of
  $D_p^b(G)$ Theories},'' \href{http://arxiv.org/abs/2106.03878}{{\ttfamily
  arXiv:2106.03878 [hep-th]}}.

\bibitem{Cvetic:2021sxm}
M.~Cvetic, M.~Dierigl, L.~Lin, and H.~Y. Zhang, ``{Higher-Form Symmetries and
  Their Anomalies in M-/F-Theory Duality},''
  \href{http://arxiv.org/abs/2106.07654}{{\ttfamily arXiv:2106.07654
  [hep-th]}}.

\bibitem{Buican:2021xhs}
M.~Buican and H.~Jiang, ``{1-Form Symmetry, Isolated N=2 SCFTs, and Calabi-Yau
  Threefolds},'' \href{http://arxiv.org/abs/2106.09807}{{\ttfamily
  arXiv:2106.09807 [hep-th]}}.

\bibitem{Bhardwaj:2021zrt}
L.~Bhardwaj, M.~Hubner, and S.~Schafer-Nameki, ``{Liberating Confinement from
  Lagrangians: 1-form Symmetries and Lines in 4d $\mathcal{N}=1$ from 6d
  $\mathcal{N}=(2,0)$},'' \href{http://arxiv.org/abs/2106.10265}{{\ttfamily
  arXiv:2106.10265 [hep-th]}}.

\bibitem{Iqbal:2021rkn}
N.~Iqbal and J.~McGreevy, ``{Mean string field theory: Landau-Ginzburg theory
  for 1-form symmetries},'' \href{http://arxiv.org/abs/2106.12610}{{\ttfamily
  arXiv:2106.12610 [hep-th]}}.

\bibitem{Braun:2021sex}
A.~P. Braun, P.-K. Oehlmann, and M.~Larfors, ``{Gauged 2-form Symmetries in 6D
  SCFTs Coupled to Gravity},''
  \href{http://arxiv.org/abs/2106.13198}{{\ttfamily arXiv:2106.13198
  [hep-th]}}.

\bibitem{Cvetic:2021maf}
M.~Cveti\v{c}, J.~J. Heckman, E.~Torres, and G.~Zoccarato, ``{Reflections on
  the Matter of 3d $\mathcal{N} = 1$ Vacua and Local $Spin(7)$
  Compactifications},'' \href{http://arxiv.org/abs/2107.00025}{{\ttfamily
  arXiv:2107.00025 [hep-th]}}.

\bibitem{Closset:2021lhd}
C.~Closset and H.~Magureanu, ``{The $U$-plane of rank-one 4d $\mathcal{N}=2$ KK
  theories},'' \href{http://arxiv.org/abs/2107.03509}{{\ttfamily
  arXiv:2107.03509 [hep-th]}}.

\bibitem{Hidaka:2021mml}
Y.~Hidaka, M.~Nitta, and R.~Yokokura, ``{Topological axion electrodynamics and
  4-group symmetry},'' \href{http://arxiv.org/abs/2107.08753}{{\ttfamily
  arXiv:2107.08753 [hep-th]}}.

\bibitem{Lee:2021obi}
Y.~Lee and Y.~Zheng, ``{Remarks on compatibility between conformal symmetry and
  continuous higher-form symmetries},''
  \href{http://dx.doi.org/10.1103/PhysRevD.104.085005}{{\em Phys. Rev. D}
  {\bfseries 104} no.~8, (2021) 085005},
  \href{http://arxiv.org/abs/2108.00732}{{\ttfamily arXiv:2108.00732
  [hep-th]}}.

\bibitem{Hidaka:2021kkf}
Y.~Hidaka, M.~Nitta, and R.~Yokokura, ``{Global 4-group symmetry and 't Hooft
  anomalies in topological axion electrodynamics},''
  \href{http://arxiv.org/abs/2108.12564}{{\ttfamily arXiv:2108.12564
  [hep-th]}}.

\bibitem{Koide:2021zxj}
M.~Koide, Y.~Nagoya, and S.~Yamaguchi, ``{Non-invertible topological defects in
  4-dimensional $\mathbb{Z}_2$ pure lattice gauge theory},''
  \href{http://arxiv.org/abs/2109.05992}{{\ttfamily arXiv:2109.05992
  [hep-th]}}.

\bibitem{Kaidi:2021xfk}
J.~Kaidi, K.~Ohmori, and Y.~Zheng, ``{Kramers-Wannier-like duality defects in
  (3 + 1)d gauge theories},'' \href{http://arxiv.org/abs/2111.01141}{{\ttfamily
  arXiv:2111.01141 [hep-th]}}.

\bibitem{Choi:2021kmx}
Y.~Choi, C.~Cordova, P.-S. Hsin, H.~T. Lam, and S.-H. Shao, ``{Non-Invertible
  Duality Defects in 3+1 Dimensions},''
  \href{http://arxiv.org/abs/2111.01139}{{\ttfamily arXiv:2111.01139
  [hep-th]}}.

\bibitem{Bah:2021brs}
I.~Bah, F.~Bonetti, E.~Leung, and P.~Weck, ``{M5-branes Probing Flux
  Backgrounds},'' \href{http://arxiv.org/abs/2111.01790}{{\ttfamily
  arXiv:2111.01790 [hep-th]}}.

\bibitem{Gukov:2021swm}
S.~Gukov, D.~Pei, C.~Reid, and A.~Shehper, ``{Symmetries of 2d TQFTs and
  Equivariant Verlinde Formulae for General Groups},''
  \href{http://arxiv.org/abs/2111.08032}{{\ttfamily arXiv:2111.08032
  [hep-th]}}.

\bibitem{Closset:2021lwy}
C.~Closset, S.~Sch\"afer-Nameki, and Y.-N. Wang, ``{Coulomb and Higgs Branches
  from Canonical Singularities, Part 1: Hypersurfaces with Smooth Calabi-Yau
  Resolutions},'' \href{http://arxiv.org/abs/2111.13564}{{\ttfamily
  arXiv:2111.13564 [hep-th]}}.

\bibitem{Yu:2021zmu}
M.~Yu, ``{Gauging Categorical Symmetries in 3d Topological Orders and Bulk
  Reconstruction},'' \href{http://arxiv.org/abs/2111.13697}{{\ttfamily
  arXiv:2111.13697 [hep-th]}}.

\bibitem{Apruzzi:2021nmk}
F.~Apruzzi, F.~Bonetti, I.~n.~G. Etxebarria, S.~S. Hosseini, and
  S.~Schafer-Nameki, ``{Symmetry TFTs from String Theory},''
  \href{http://arxiv.org/abs/2112.02092}{{\ttfamily arXiv:2112.02092
  [hep-th]}}.

\bibitem{Beratto:2021xmn}
E.~Beratto, N.~Mekareeya, and M.~Sacchi, ``{Zero-form and one-form symmetries
  of the ABJ and related theories},''
  \href{http://arxiv.org/abs/2112.09531}{{\ttfamily arXiv:2112.09531
  [hep-th]}}.

\bibitem{Bhardwaj:2021mzl}
L.~Bhardwaj, S.~Giacomelli, M.~H\"ubner, and S.~Sch\"afer-Nameki, ``{Relative
  Defects in Relative Theories: Trapped Higher-Form Symmetries and Irregular
  Punctures in Class S},'' \href{http://arxiv.org/abs/2201.00018}{{\ttfamily
  arXiv:2201.00018 [hep-th]}}.

\bibitem{DelZotto:2022fnw}
M.~Del~Zotto, J.~J. Heckman, S.~N. Meynet, R.~Moscrop, and H.~Y. Zhang,
  ``{Higher Symmetries of 5d Orbifold SCFTs},''
  \href{http://arxiv.org/abs/2201.08372}{{\ttfamily arXiv:2201.08372
  [hep-th]}}.

\bibitem{Chatterjee:2022kxb}
A.~Chatterjee and X.-G. Wen, ``{Algebra of local symmetric operators and
  braided fusion $n$-category -- symmetry is a shadow of topological order},''
  \href{http://arxiv.org/abs/2203.03596}{{\ttfamily arXiv:2203.03596
  [cond-mat.str-el]}}.

\bibitem{Benini:2022hzx}
F.~Benini, C.~Copetti, and L.~Di~Pietro, ``{Factorization and Global Symmetries
  in Holography},'' \href{http://arxiv.org/abs/2203.09537}{{\ttfamily
  arXiv:2203.09537 [hep-th]}}.

\bibitem{Apruzzi:2022dlm}
F.~Apruzzi, ``{Higher Form Symmetries TFT in 6d},''
  \href{http://arxiv.org/abs/2203.10063}{{\ttfamily arXiv:2203.10063
  [hep-th]}}.

\bibitem{Hubner:2022kxr}
M.~Hubner, D.~R. Morrison, S.~Schafer-Nameki, and Y.-N. Wang, ``{Generalized
  Symmetries in F-theory and the Topology of Elliptic Fibrations},''
  \href{http://arxiv.org/abs/2203.10022}{{\ttfamily arXiv:2203.10022
  [hep-th]}}.

\bibitem{Lee:2022spd}
Y.~Lee and K.~Yonekura, ``{Global anomalies in 8d supergravity},''
  \href{http://arxiv.org/abs/2203.12631}{{\ttfamily arXiv:2203.12631
  [hep-th]}}.

\bibitem{Carta:2022spy}
F.~Carta, S.~Giacomelli, N.~Mekareeya, and A.~Mininno, ``{Dynamical
  consequences of 1-form symmetries and the exceptional Argyres-Douglas
  theories},'' \href{http://arxiv.org/abs/2203.16550}{{\ttfamily
  arXiv:2203.16550 [hep-th]}}.

\bibitem{Roumpedakis:2022aik}
K.~Roumpedakis, S.~Seifnashri, and S.-H. Shao, ``{Higher Gauging and
  Non-invertible Condensation Defects},''
  \href{http://arxiv.org/abs/2204.02407}{{\ttfamily arXiv:2204.02407
  [hep-th]}}.

\bibitem{Bhardwaj:2022yxj}
L.~Bhardwaj, L.~Bottini, S.~Schafer-Nameki, and A.~Tiwari, ``{Non-Invertible
  Higher-Categorical Symmetries},''
  \href{http://arxiv.org/abs/2204.06564}{{\ttfamily arXiv:2204.06564
  [hep-th]}}.

\bibitem{DelZotto:2022ras}
M.~Del~Zotto and I.~n. Etxebarria~Garc\'\i{}a, ``{Global Structures from the
  Infrared},'' \href{http://arxiv.org/abs/2204.06495}{{\ttfamily
  arXiv:2204.06495 [hep-th]}}.

\bibitem{Hayashi:2022fkw}
Y.~Hayashi and Y.~Tanizaki, ``{Non-invertible self-duality defects of
  Cardy-Rabinovici model and mixed gravitational anomaly},''
  \href{http://arxiv.org/abs/2204.07440}{{\ttfamily arXiv:2204.07440
  [hep-th]}}.

\bibitem{Choi:2022zal}
Y.~Choi, C.~Cordova, P.-S. Hsin, H.~T. Lam, and S.-H. Shao, ``{Non-invertible
  Condensation, Duality, and Triality Defects in 3+1 Dimensions},''
  \href{http://arxiv.org/abs/2204.09025}{{\ttfamily arXiv:2204.09025
  [hep-th]}}.

\bibitem{Argyres:2022kon}
P.~C. Argyres, M.~Martone, and M.~Ray, ``{Dirac pairings, one-form symmetries
  and Seiberg-Witten geometries},''
  \href{http://arxiv.org/abs/2204.09682}{{\ttfamily arXiv:2204.09682
  [hep-th]}}.

\bibitem{Kaidi:2022uux}
J.~Kaidi, G.~Zafrir, and Y.~Zheng, ``{Non-Invertible Symmetries of
  $\mathcal{N}=4$ SYM and Twisted Compactification},''
  \href{http://arxiv.org/abs/2205.01104}{{\ttfamily arXiv:2205.01104
  [hep-th]}}.

\bibitem{Heckman:2022suy}
J.~J. Heckman, C.~Lawrie, L.~Lin, H.~Y. Zhang, and G.~Zoccarato, ``{6d SCFTs,
  Center-Flavor Symmetries, and Stiefel--Whitney Compactifications},''
  \href{http://arxiv.org/abs/2205.03411}{{\ttfamily arXiv:2205.03411
  [hep-th]}}.

\bibitem{Benedetti:2022zbb}
V.~Benedetti, H.~Casini, and J.~M. Magan, ``{Generalized symmetries and
  Noether's theorem in QFT},''
  \href{http://arxiv.org/abs/2205.03412}{{\ttfamily arXiv:2205.03412
  [hep-th]}}.

\bibitem{Choi:2022jqy}
Y.~Choi, H.~T. Lam, and S.-H. Shao, ``{Non-invertible Global Symmetries in the
  Standard Model},'' \href{http://arxiv.org/abs/2205.05086}{{\ttfamily
  arXiv:2205.05086 [hep-th]}}.

\bibitem{Cordova:2022ieu}
C.~Cordova and K.~Ohmori, ``{Non-Invertible Chiral Symmetry and Exponential
  Hierarchies},'' \href{http://arxiv.org/abs/2205.06243}{{\ttfamily
  arXiv:2205.06243 [hep-th]}}.

\bibitem{Chatterjee:2022tyg}
A.~Chatterjee and X.-G. Wen, ``{Holographic theory for the emergence and the
  symmetry protection of gaplessness and for continuous phase transitions},''
  \href{http://arxiv.org/abs/2205.06244}{{\ttfamily arXiv:2205.06244
  [cond-mat.str-el]}}.

\bibitem{Lohitsiri:2022jyz}
N.~Lohitsiri and T.~Sulejmanpasic, ``{Comments on QCD$_3$ and anomalies with
  fundamental and adjoint matter},''
  \href{http://arxiv.org/abs/2205.07825}{{\ttfamily arXiv:2205.07825
  [hep-th]}}.

\bibitem{Pantev:2022kpl}
T.~Pantev, D.~Robbins, E.~Sharpe, and T.~Vandermeulen, ``{Orbifolds by 2-groups
  and decomposition},'' \href{http://arxiv.org/abs/2204.13708}{{\ttfamily
  arXiv:2204.13708 [hep-th]}}.

\bibitem{Sharpe:2021srf}
E.~Sharpe, ``{Topological operators, noninvertible symmetries and
  decomposition},'' \href{http://arxiv.org/abs/2108.13423}{{\ttfamily
  arXiv:2108.13423 [hep-th]}}.

\bibitem{Robbins:2021xce}
D.~G. Robbins, E.~Sharpe, and T.~Vandermeulen, ``{Anomaly resolution via
  decomposition},'' \href{http://dx.doi.org/10.1142/S0217751X21502201}{{\em
  Int. J. Mod. Phys. A} {\bfseries 36} no.~29, (2021) 2150220},
  \href{http://arxiv.org/abs/2107.13552}{{\ttfamily arXiv:2107.13552
  [hep-th]}}.

\bibitem{Bhardwaj:2022dyt}
L.~Bhardwaj, M.~Bullimore, A.~E.~V. Ferrari, and S.~Schafer-Nameki,
  ``{Anomalies of Generalized Symmetries from Solitonic Defects},''
  \href{http://arxiv.org/abs/2205.15330}{{\ttfamily arXiv:2205.15330
  [hep-th]}}.

\bibitem{Bolognesi:2022beq}
S.~Bolognesi, K.~Konishi, and A.~Luzio, ``{Dynamical Abelianization and
  anomalies in chiral gauge theories},''
  \href{http://arxiv.org/abs/2206.00538}{{\ttfamily arXiv:2206.00538
  [hep-th]}}.

\bibitem{Argyres:2016yzz}
P.~C. Argyres and M.~Martone, ``{4d $ \mathcal{N} $ =2 theories with
  disconnected gauge groups},''
  \href{http://dx.doi.org/10.1007/JHEP03(2017)145}{{\em JHEP} {\bfseries 03}
  (2017) 145}, \href{http://arxiv.org/abs/1611.08602}{{\ttfamily
  arXiv:1611.08602 [hep-th]}}.

\bibitem{Cordova:2017vab}
C.~Cordova, P.-S. Hsin, and N.~Seiberg, ``{Global Symmetries, Counterterms, and
  Duality in Chern-Simons Matter Theories with Orthogonal Gauge Groups},''
  \href{http://dx.doi.org/10.21468/SciPostPhys.4.4.021}{{\em SciPost Phys.}
  {\bfseries 4} no.~4, (2018) 021},
  \href{http://arxiv.org/abs/1711.10008}{{\ttfamily arXiv:1711.10008
  [hep-th]}}.

\bibitem{Bourget:2018ond}
A.~Bourget, A.~Pini, and D.~Rodr\'\i{}guez-G\'omez, ``{Gauge theories from
  principally extended disconnected gauge groups},''
  \href{http://dx.doi.org/10.1016/j.nuclphysb.2019.02.004}{{\em Nucl. Phys. B}
  {\bfseries 940} (2019) 351--376},
  \href{http://arxiv.org/abs/1804.01108}{{\ttfamily arXiv:1804.01108
  [hep-th]}}.

\bibitem{Arias-Tamargo:2019jyh}
G.~Arias-Tamargo, A.~Bourget, A.~Pini, and D.~Rodr\'\i{}guez-G\'omez,
  ``{Discrete gauge theories of charge conjugation},''
  \href{http://dx.doi.org/10.1016/j.nuclphysb.2019.114721}{{\em Nucl. Phys. B}
  {\bfseries 946} (2019) 114721},
  \href{http://arxiv.org/abs/1903.06662}{{\ttfamily arXiv:1903.06662
  [hep-th]}}.

\bibitem{Arias-Tamargo:2021ppf}
G.~Arias-Tamargo, A.~Bourget, and A.~Pini, ``{Discrete gauging and Hasse
  diagrams},'' \href{http://dx.doi.org/10.21468/SciPostPhys.11.2.026}{{\em
  SciPost Phys.} {\bfseries 11} no.~2, (2021) 026},
  \href{http://arxiv.org/abs/2105.08755}{{\ttfamily arXiv:2105.08755
  [hep-th]}}.

\bibitem{Henning:2021ctv}
B.~Henning, X.~Lu, T.~Melia, and H.~Murayama, ``{Outer Automorphism
  Anomalies},'' \href{http://arxiv.org/abs/2111.04728}{{\ttfamily
  arXiv:2111.04728 [hep-th]}}.

\end{thebibliography}\endgroup

\end{document}